\documentclass[a4paper,12pt]{article}

\usepackage[left=2.5cm,right=2.5cm,top=2.5cm,bottom=2.5cm]{geometry}
\usepackage{graphicx}
\usepackage{amssymb}
\usepackage{multirow}
\usepackage{amsmath}
\usepackage{psfrag}
\usepackage{setspace}
\usepackage{subcaption}
\usepackage[colorlinks=true,bookmarks=true]{hyperref}
\usepackage{float}

\hypersetup{citecolor=blue}
\newcommand{\dif}{\mathrm{d}}

\newcommand{\Eqref}[1]{(\ref{#1})}
\newcommand{\half}{\frac{1}{2}}

\newcommand{\brac}[1]{\left(#1 \right)}
\newcommand{\sbrac}[1]{\left[#1\right]}

\begin{document}

\title{Motion of charged particles around a magnetized/electrified black hole}
\author{Yen-Kheng Lim\footnote{E-mail: phylyk@nus.edu.sg}\\\textit{Department of Physics, National University of Singapore,}\\\textit{Singapore 117551, Singapore}}
\date{\today}
\maketitle


\begin{abstract}
 Geodesic equations of timelike and null charged particles in the Ernst metric are studied. We consider two distinct forms of the Ernst solution where the Maxwell potential represents either a uniform electric or magnetic field. Circular orbits in various configurations are considered, as well as their perturbations and stability. We find that the electric field strength must be below a certain charge-dependent critical value for these orbits to be stable. The case of the magnetic Ernst metric contains a limit which reduces to the Melvin magnetic universe. In this case the equations of motion are solved to reveal cycloidlike or trochoidlike motion, similar to those found by Frolov and Shoom around black holes immersed in test magnetic fields.
\end{abstract}

\newpage

\section{Introduction} 

The Ernst metric \cite{Ernst:1975}, also known as the Schwarzschild-Melvin metric or the electrified/magnetized Schwarzschild black hole, describes a black hole immersed in an uniform electric or magnetic field. In this paper we will distinguish the two cases by calling the former the \emph{electric} Ernst spacetime and the latter \emph{magnetic} Ernst spacetime. The solution can be derived by applying a Harrison transformation \cite{Harrison:1968} to a seed metric taken to be the Schwarzschild solution. The result is an exact solution to the Einstein-Maxwell equations parametrized by the black hole mass and the strength of the Maxwell field. The spacetime curvature is caused by both the black hole and the electromagnetic field. The Ernst solution was found just a year after Wald considered Kerr black holes immersed in a weak magnetic field \cite{PhysRevD.10.1680}. In Wald's solution a (test) magnetic field is found by solving Maxwell's equation in a Kerr background. Hence the field in Wald's construction is a test field which does not influence the spacetime curvature and can be considered as a limiting special case of the full Ernst solution.
\par 
Despite being non-asymptotically flat, the magnetic Ernst solution is a useful model which describes black holes in certain astrophysical situations. Hence past work on geodesics focuses primarily on the magnetic case \cite{2013Natur.501..391E,Kardashev:1995}. It is likely that strong magnetic fields may exist in the vicinity of stellar mass and supermassive black holes. In particular, the motion of charged particles in this spacetime is studied in \cite{Dadhich1979}. The geodesic motion for null and timelike particles is considered in \cite{Dhurandhar:1983,Estaban1984,Esteban:1985}. Some later works on this topic include \cite{Bakhankov:1989,Karas1990,Brito:2014nja}. The motion of charged particles was also considered in the context of chaos and nonlinear dynamics in \cite{Karas1992,Santoprete:2001}.
\par 
Nevertheless, in most realistic astrophysical situations the magnetic field surrounding a black hole is relatively weak such that the influence on the spacetime curvature is fairly negligible \cite{Aliev:1989}. The motion of charged particles in this test field regime was studied in Refs.~\cite{Perti:2004,Aliev:2002nw,Frolov:2010mi,Frolov:2014zia}, where the electromagnetic field does not influence the spacetime curvature and only affects motion of the charged particle via Lorentz interactions. Frolov and Shoom \cite{Frolov:2010mi}, in particular, studied the case where the particle executes a curly cycloidlike trajectory which occurs due to the combination of a central gravitational force from the black hole on the particle together with a velocity-dependent outward Lorentz force. In this paper, we will also consider a similar motion in the Melvin spacetime, where, in the absence of a black hole, the inward gravitational force is provided by the magnetic field itself.
\par 
The case of orbits around a black hole immersed in an electric field has received less attention compared to its magnetic counterpart. This case physically describes an electrically charged particle under the gravitational influence of the black hole and the axisymmetric electric field, in addition to a constant Coulomb force in the $z$ direction. This can be viewed physically as a central force motion subject to an additional uniform constant force, we should expect some similarities to particle motion in the vacuum C-metric \cite{Lim:2014qra}.
\par 
This paper is organised as follows: In Sec.~\ref{metric} we will review the Ernst metric and some of its relevant properties. In Sec.~\ref{ernstE} we derive the equations of motion for a charged particle in the electric Ernst spacetime and study some solutions which are accessible analytically and perturbatively. Section \ref{ernstM} follows by deriving the equations of motion for the case of charged particle in the magnetic Ernst spacetime. For neutral particles the equations of motion for the electric and magnetic cases are identical. This includes the motion of null (lightlike) particles. This case has already been thoroughly studied in Refs.~\cite{Dhurandhar:1983,Estaban1984,Esteban:1985} using the Hamilton-Jacobi formalism. For completeness we consider them in Sec.~\ref{neutral} using the Lagrangian formalism we adopt for this paper. Some concluding remarks are discussed in Sec.~\ref{conclusion}.

\section{Ernst spacetime}\label{metric}

The four-dimensional Ernst metric \cite{Ernst:1975} is given by\footnote{A review and related references can be found in \cite{Griffiths:2009dfa}.}
\begin{align}
 \dif s^2=&\;\Lambda^2\brac{-f\dif t^2+f^{-1}\dif r^2+r^2\dif\theta^2}+\Lambda^{-2}r^2\sin^2\theta\,\dif\phi^2,\nonumber\\
     f=&\;1-\frac{2m}{r},\quad \Lambda=1+\frac{1}{4}B^2r^2\sin^2\theta, \label{Ernst_metric}
\end{align}
where $m$ parametrizes the mass of the black hole and $B$ parametrizes the strength of the uniform electric/magnetic field.\footnote{We will keep the notation where $B$ represents the field strength parameter for both the electric and magnetic Ernst solutions, as they will be discussed separately in this paper. Thus it will be clear from the context that $B$ represents the \emph{electric} field strength for the electric Ernst solution (Sec.~\ref{ernstE}) while it represents the \emph{magnetic} field strength in the analysis of the magnetic Ernst solution (Sec.~\ref{ernstM}).} If the black hole is immersed in a magnetic field, the Maxwell potential is
\begin{align}
 A_{\mathrm{M}}=&\;B\frac{r^2\sin^2\theta}{2\Lambda}\,\dif\phi. \label{A_M}
\end{align}
On the other hand, the Maxwell potential corresponding to an electrified black hole is
\begin{align}
 A_{\mathrm{E}}=&\;Bfr\cos\theta\,\dif t. \label{A_E}
\end{align}
It is worth noting that with the presence of the electromagnetic field, the Ernst solution is not asymptotically flat. The Maxwell tensor for either case is given by $F_{\mu\nu}=\partial_\mu A_\nu-\partial_\nu A_\mu$. The metric \Eqref{Ernst_metric}, together with either \Eqref{A_M} or \Eqref{A_E} is a solution to the four-dimensional Einstein-Maxwell equations with zero cosmological constant. We can easily see that setting $m=0$ reduces the solution to that of Melvin \cite{Melvin:1965zza}, while for $B=0$ we recover the usual Schwarzschild solution. 
\par 
We also note that there are some useful symmetries of the metric. For the case of the magnetic Ernst spacetime, \Eqref{Ernst_metric} and \Eqref{A_M} are invariant under the transformation
\begin{align}
 B\rightarrow -B,\quad\phi\rightarrow-\phi. \label{ernstM_sym}
\end{align}
For the case of the electric Ernst spacetime, the solutions \Eqref{Ernst_metric} and \Eqref{A_E} are invariant under
\begin{align}
 B\rightarrow -B,\quad\theta\rightarrow\pi-\theta. \label{ernstE_sym}
\end{align}
With these symmetries, we can consider only the case $B>0$ without loss of generality.

\section{Charged particles in the electric Ernst spacetime} \label{ernstE}

\subsection{Equations of motion}

The motion of a test particle of charge per unit mass $e$ is described by a trajectory $x^\mu(\tau)$ where $\tau$ is an appropriate affine parametrization. In the case of timelike particles, $\tau$ may be regarded as the proper time measured by the particle. The motion is determined by the Lagrangian $\mathcal{L}=\half g_{\mu\nu}\dot{x}^\mu\dot{x}^\nu+eA_\mu\dot{x}^\mu$, where overdots denote derivatives with respect to $\tau$. The subsequent equations of motion can be derived using the Euler-Lagrange equation $\frac{\dif}{\dif\tau}\frac{\partial\mathcal{L}}{\partial\dot{x}^\mu}=\frac{\partial\mathcal{L}}{\partial x^\mu}$. For the case of charged particles in the electric Ernst spacetime, where the Maxwell potential is given by Eq.~\Eqref{A_E},  the corresponding Lagrangian is
\begin{align}
 \mathcal{L}=&\;\half\sbrac{\Lambda^2\brac{-f\dot{t}^2+\frac{\dot{r}^2}{f}+r^2\dot{\theta}^2}+\frac{r^2\sin^2\theta}{\Lambda^2}\dot{\phi}^2}+eBfr\cos\theta\,\dot{t}.
\end{align}
Since $\partial/\partial t$ and $\partial/\partial\phi$ are Killing vectors, they give rise to constants of motion $E$ and $\Phi$, which we may interpret as the energy and angular momentum of the particle. The conserved quantities reduce the equations for $t$ and $\phi$ into first integrals:
\begin{align}
 \dot{t}=\frac{E+eBr\cos\theta}{\Lambda^2f},\quad\dot{\phi}=\frac{\Lambda^2\Phi}{r^2\sin^2\theta}. \label{ernstE_cons}
\end{align}
Applying the Euler-Lagrange equation to the remaining two coordinates gives
\begin{align}
 \ddot{r}=&\;\brac{\frac{f'}{2f}-\frac{\partial_r\Lambda}{\Lambda}}\dot{r}^2+f\brac{r+\frac{r^2\partial_r\Lambda}{\Lambda}}\dot{\theta}^2-\frac{2\partial_\theta\Lambda}{\Lambda}\dot{r}\dot{\theta}+f\brac{1-\frac{r\partial_r\Lambda}{\Lambda}}\frac{\Phi^2}{r^3\sin^2\theta}\nonumber\\
     &\;-\frac{1}{\Lambda^4}\brac{\frac{\partial_r\Lambda}{\Lambda}+\frac{f'}{2f}}\brac{E+eBr\cos\theta}^2+\frac{eB\cos\theta}{\Lambda^4}\brac{E+eBr\cos\theta},\label{ernstE_rddot}\\
 \ddot{\theta}=&\;\frac{\partial_\theta\Lambda}{\Lambda}\brac{\frac{\dot{r}^2}{r^2f}-\dot{\theta}^2}-2\brac{1+\frac{r\partial_r\Lambda}{\Lambda}}\dot{r}\dot{\theta}+\brac{\cos\theta-\frac{\sin\theta\partial_\theta\Lambda}{\Lambda}}\frac{\Phi^2}{r^4\sin^3\theta}\nonumber\\
              &\;-\frac{\partial_\theta\Lambda}{r^2\Lambda^5f}\brac{E+eBr\cos\theta}^2-\frac{eB\sin\theta}{r\Lambda^4f}\brac{E+eBr\cos\theta}. \label{ernstE_thetaddot}
\end{align}
Here, the primes appearing in $f'$ indicate derivatives with respect to $r$. The invariance of $g_{\mu\nu}\dot{x}^\mu\dot{x}^\nu\equiv\epsilon$, together with \Eqref{ernstE_cons}, gives a first integral equation
\begin{align}
 -\frac{\brac{E+eBr\cos\theta}^2}{\Lambda^2f}+\Lambda^2\brac{\frac{\Phi^2}{r^2\sin^2\theta}+\frac{\dot{r}^2}{f}+r^2\dot{\theta}^2}=\epsilon. \label{ernstE_first}
\end{align}
By appropriately rescaling the affine parameter $\tau$, we can set the magnitude of $\epsilon$ to be unity if it is nonzero. Therefore, for timelike particles we have $\epsilon=-1$, and $\epsilon=0$ for null (massless) particles. 
\par 
We note that due to the symmetry of the solution shown in \Eqref{ernstE_sym}, we may assume without loss of generality that both $B>0$ and $e>0$. Since if both are negative the above equations remain unchanged, if they have opposite signs, the resulting equations are equivalent under \Eqref{ernstE_sym}.
\par 
If we set $m=0$ in the above equations, they reduce to the geodesic equations of the electric Melvin spacetime, while setting $B=0$ reduces to the well-known geodesic equations around a Schwarzschild black hole. Furthermore, setting $B=b/e$ and neglecting terms of order $1/e^2$ and beyond describes the motion around a black hole immersed in a test electric field. A simpler, though slightly less rigorous way to describe this is obtained by setting $\Lambda=1$ in the above equations. In the test field regime, the electric field does not influence the spacetime curvature, hence its effect on the particle is purely electrodynamic in nature.

\subsection{Curves of zero velocity}

In the present case where the black hole is immersed in an electric field, it is not possible to cast Eq.~\Eqref{ernstE_first} in the form of of an effective potential equation. Nevertheless, it is still possible to study the existence of bound and unbound orbits by rearranging to obtain
\begin{align}
 \Lambda^4\brac{\dot{r}^2+r^2f\dot{\theta}^2}=\brac{E+eBr\cos\theta}^2-\frac{\Lambda^4f\Phi^2}{r^2\sin^2\theta}+\Lambda^2f\epsilon. \label{ernstE_Veff}
\end{align}
From the above equation, we can find regions accessible to the particle where the coordinates $r$ and $\theta$ lead to positive values in the right-hand side of Eq.~\Eqref{ernstE_Veff}. The boundaries of these regions are points where $\dot{r}=\dot{\theta}=0$, and therefore represent turning points of the trajectory in which the particle reaches zero velocity. Hence, in earlier works such as  Refs.~\cite{Contopoulos:1991a,Contopoulos:1991b} call such boundaries \emph{curves of zero velocity}.
\par
\begin{figure}
 \begin{center}
  \includegraphics[scale=0.34]{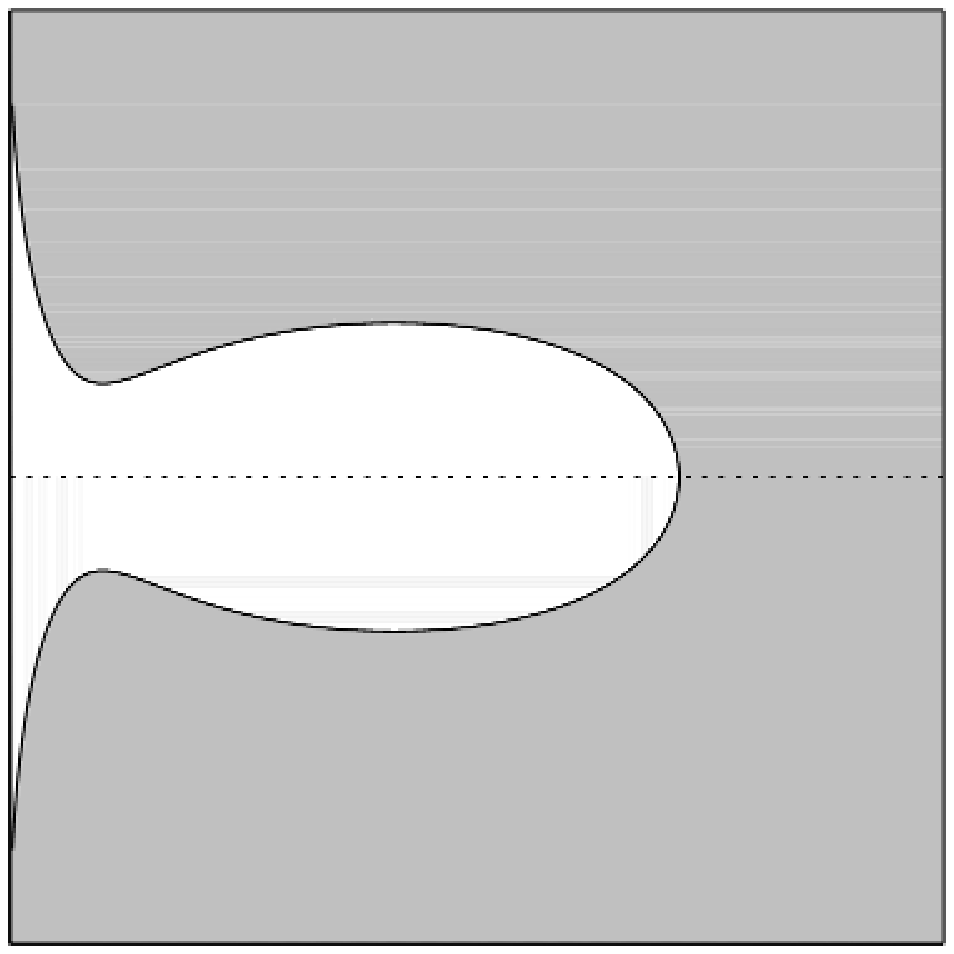}
  \includegraphics[scale=0.34]{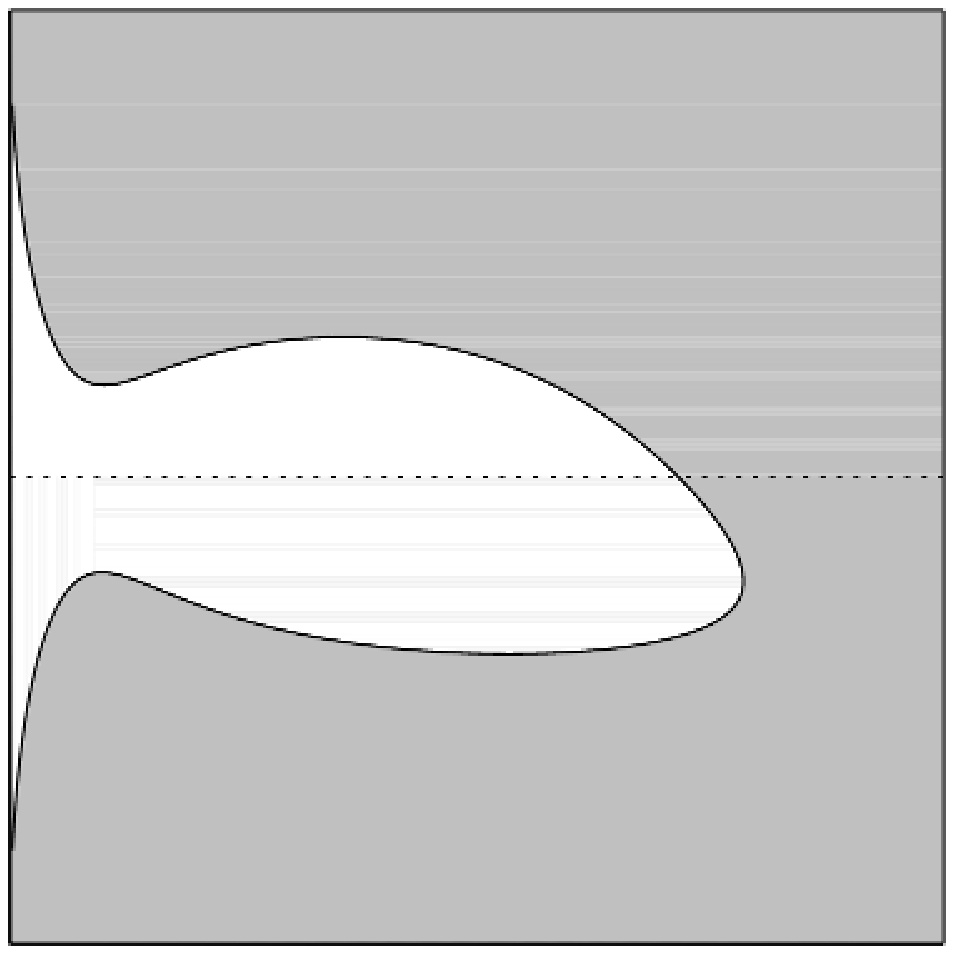}
  \includegraphics[scale=0.34]{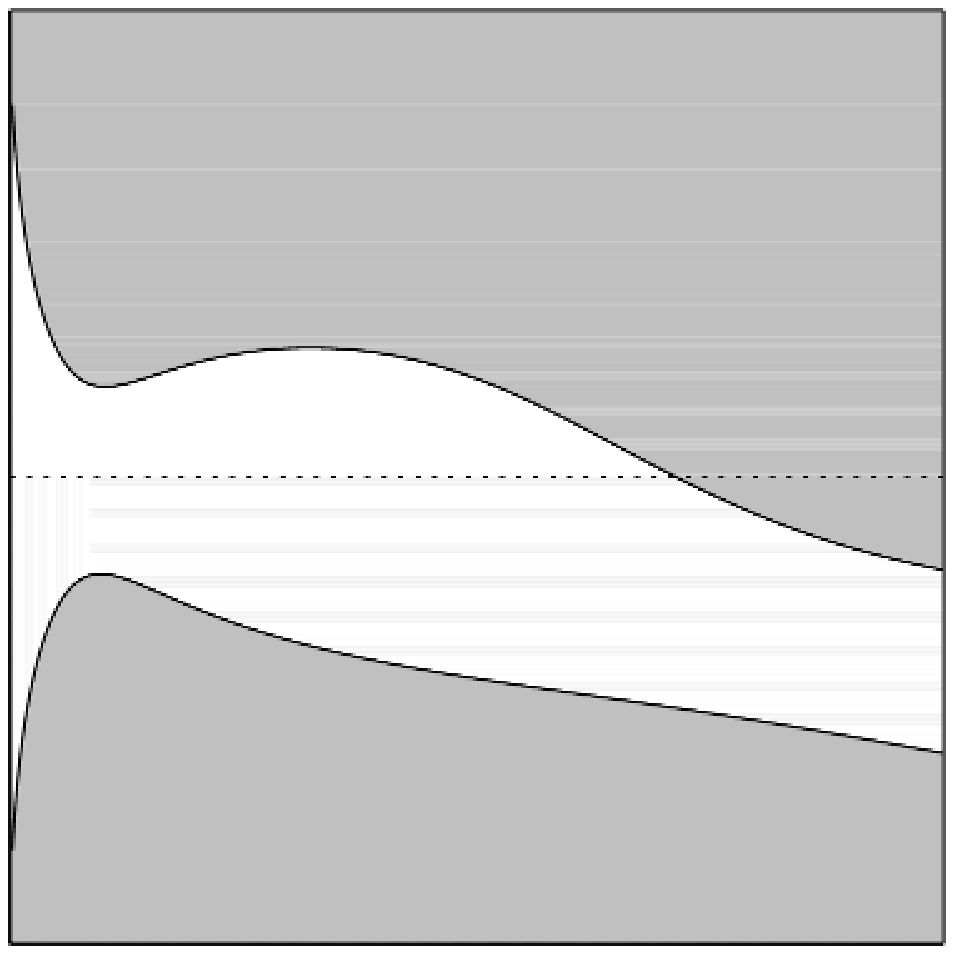}
  \includegraphics[scale=0.34]{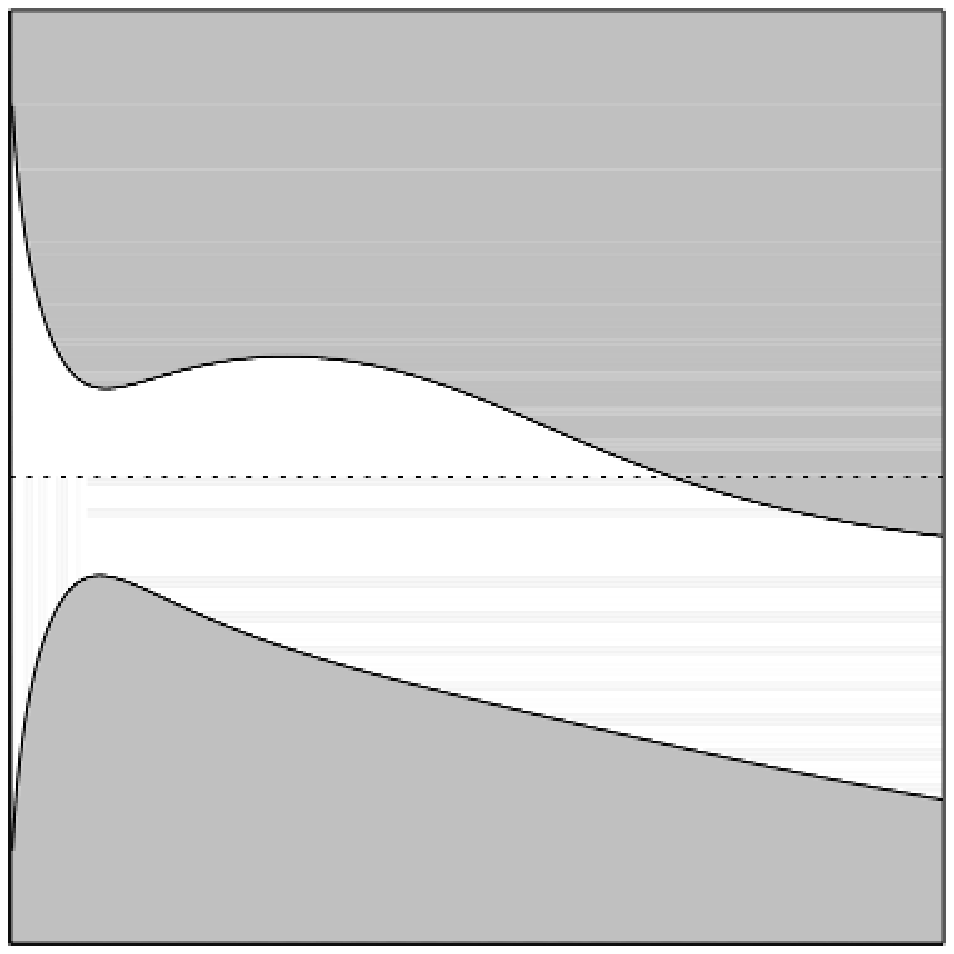}
  \\
  \includegraphics[scale=0.34]{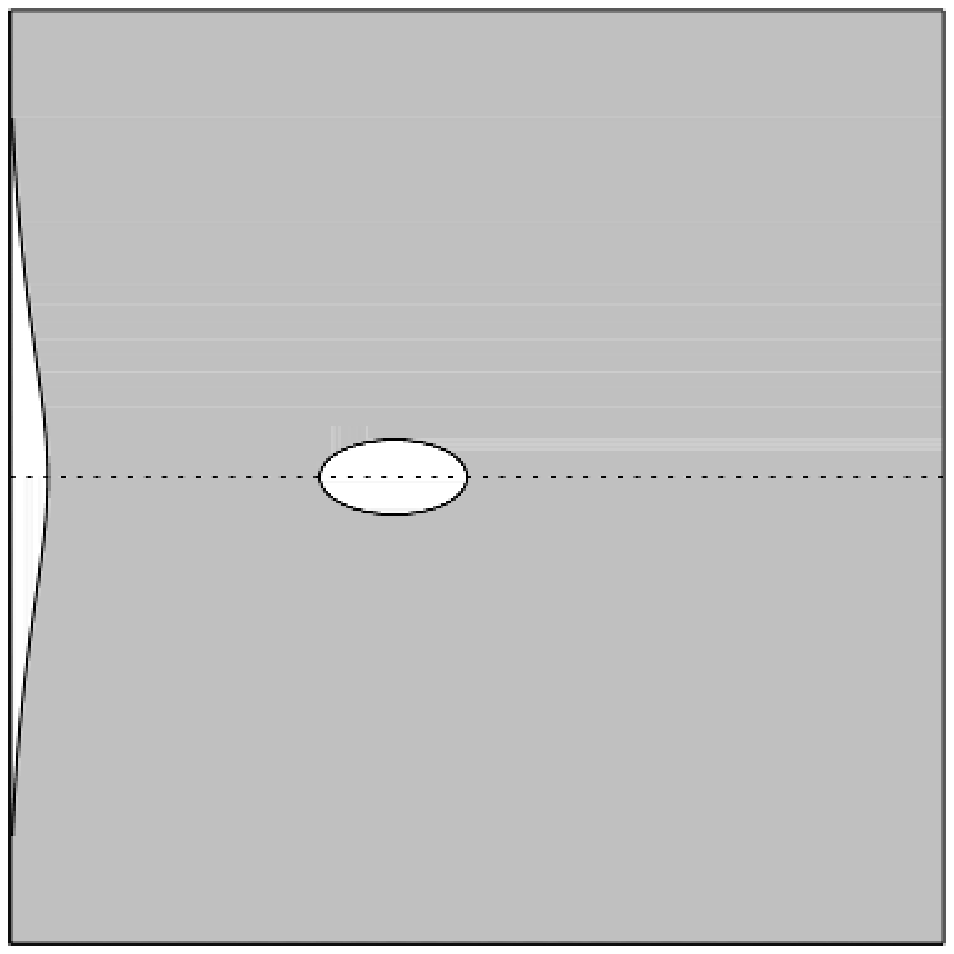}
  \includegraphics[scale=0.34]{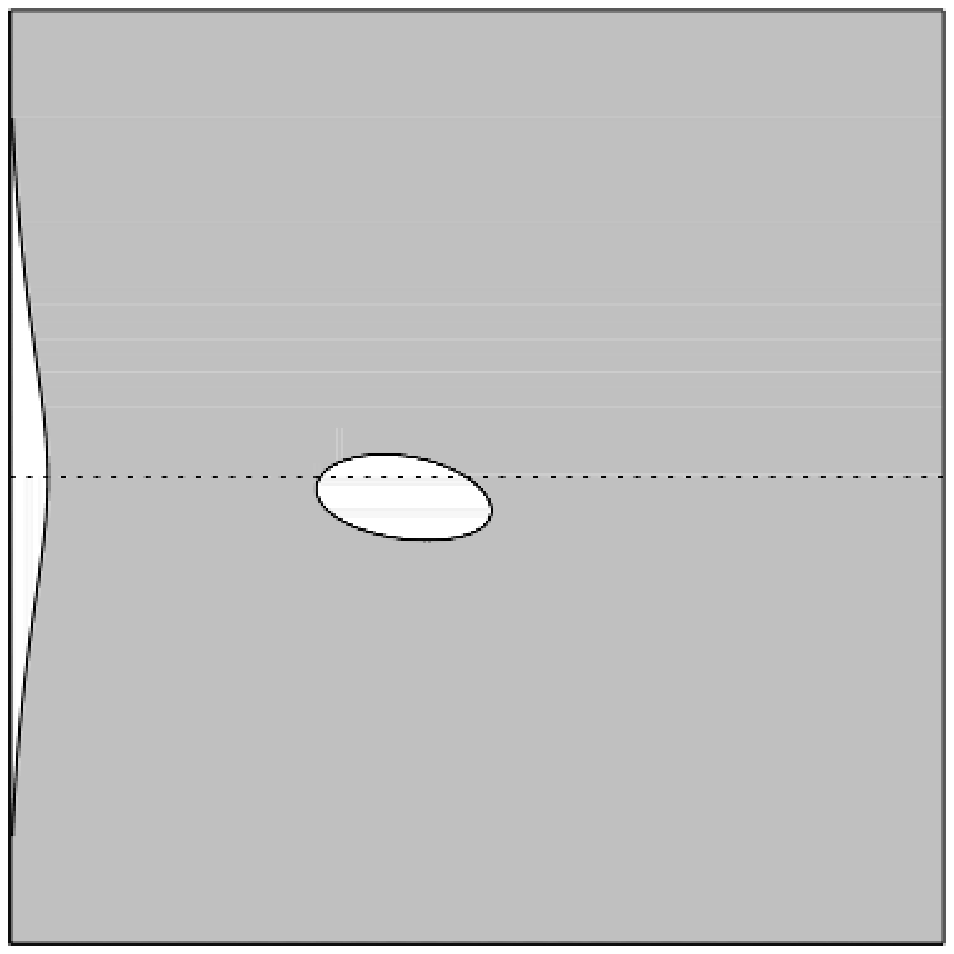}
  \includegraphics[scale=0.34]{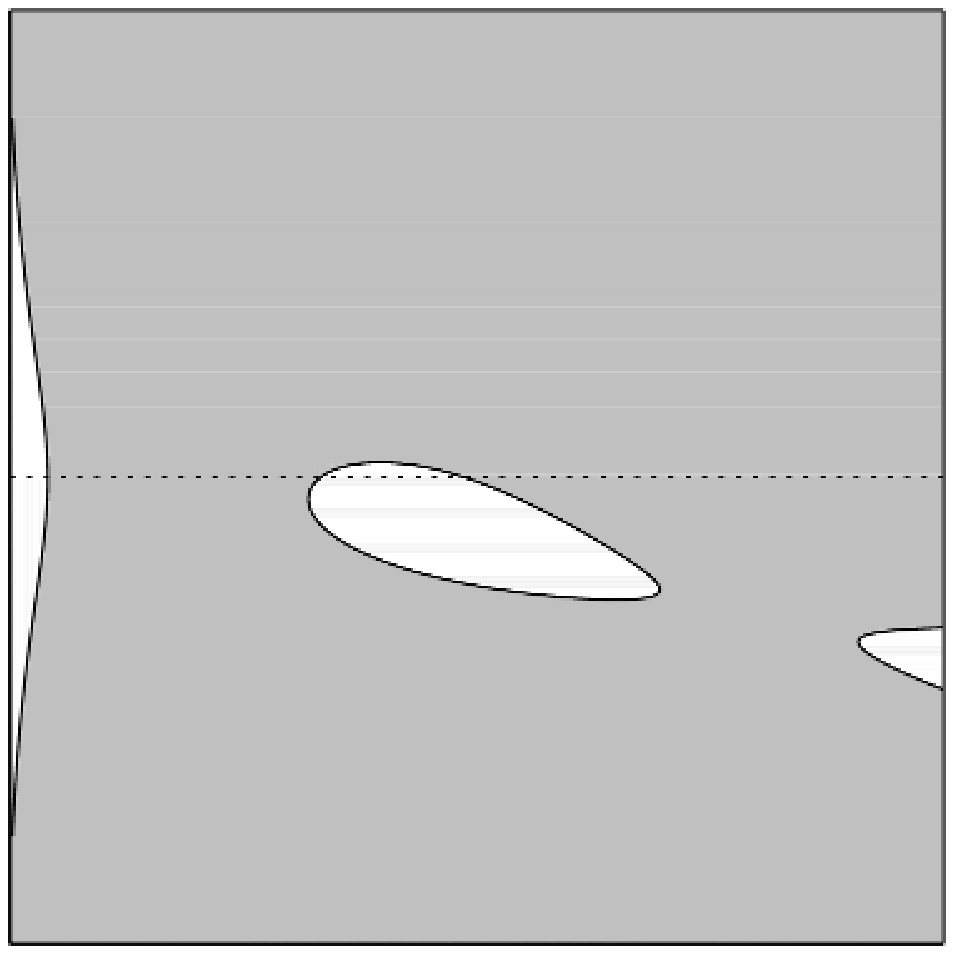}
  \includegraphics[scale=0.34]{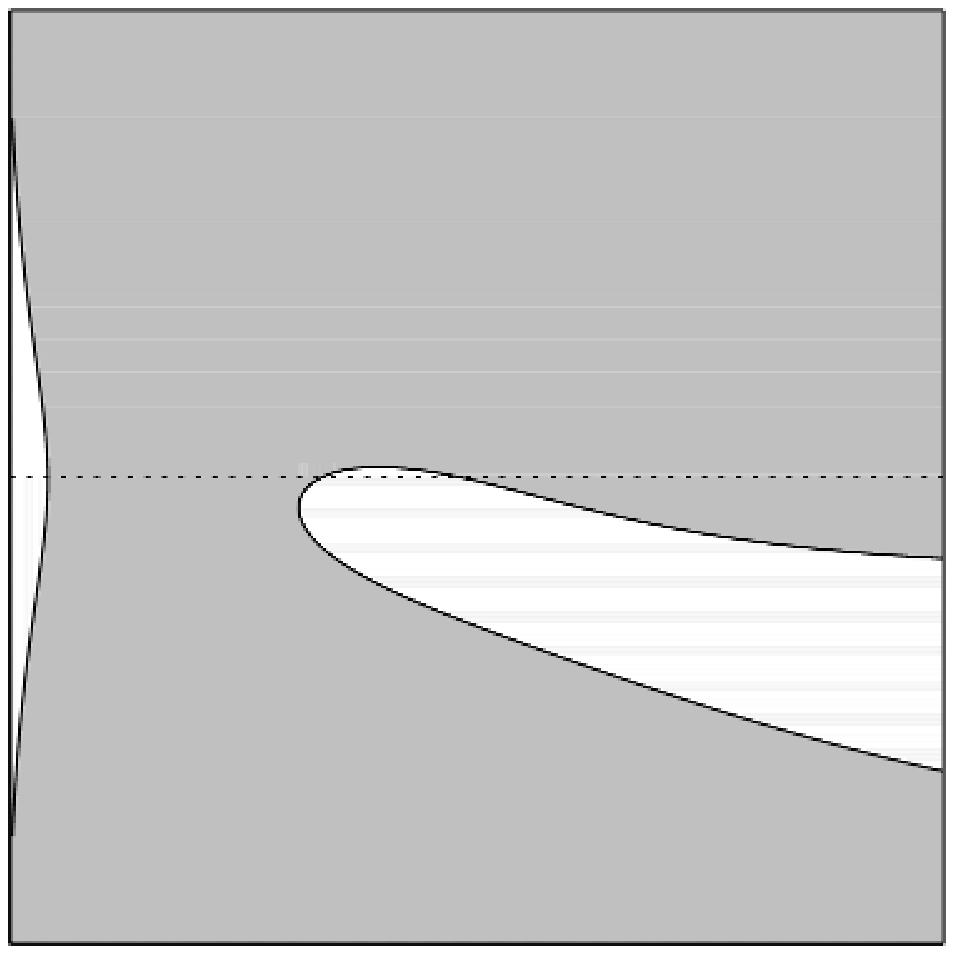}
 \end{center}
 \caption{(Color online) The accessible regions for charged particles in the electric Ernst spacetime of charge $e=1$, $E=0.963$. The vertical and horizontal axes correspond to $\theta$ and $r$, with the shaded regions as those inaccessible to the particle. The top and bottom row correspond to particles of angular momentum $\Phi=3.5$ and $4.0$, respectively. In each row, from left to right the electric field strength is $B=0$, 0.0005, 0.0010, and 0.0015. The above diagrams are plotted for the range $r\in[2m,27m]$ and $\theta\in[0,\pi]$. The equator $\theta=\pi/2$ is represented in each diagram with the horizontal dotted line.}
 \label{EV_eff}

\end{figure}

 
Figure \ref{EV_eff} shows some typical examples of the regions accessible by the charged particle. The $r$ and $\theta$ coordinates are represented by the horizontal and vertical directions, respectively. The shaded areas represents the areas inaccessible to the particle. The top row is for the case $\Phi=3.5$, while the bottom row are plots for $\Phi=4$. The leftmost diagrams on both rows show the case corresponding to $B=0$, which is simply the Schwarzschild effective potential. As can be shown from the well-known Schwarzschild geodesics, around the value of $\Phi=4$, the Schwarzschild potential contains a finite well, in which the particle is in a bound orbit around the black hole \cite{Carroll:2004st,Chandrasekhar:1985kt}. When $\Phi$ is reduced to about to 3.5, the potential barrier disappears and the particle may fall into the horizon at $r=2m$.
\par 
The electric field is turned on and increased in the second, third and fourth plots of each row. As we see by visual inspection, the accessible regions are mostly pushed below $\theta<\pi/2$ (towards the north of the equator). This corresponds to the intuitive interpretation that the north-pointing electric field tends to push the positively charged particle along that direction. If the particle is negatively charged, or if the field points in the opposite direction, the particle will be pushed southwards, and the above diagrams should be reflected along the $\theta=\pi/2$ axis where the equations of motion are equivalent under \Eqref{ernstE_sym}.

\subsection{Nearly circular orbits around weakly electrified black holes}

When $B$ is small, the spacetime can be regarded as a perturbation of the Schwarzschild solution. In this subsection we shall consider what happens to circular Schwarzschild orbits in the presence of a weak electric field. 
\par 
We first consider the case where the electric field is parallel to the orbital plane; hence, in our coordinate system this is a polar orbit, where $\Phi=0$, and now we only have angular momentum in the $\theta$ direction, which we denote as $L$. Recall that circular Schwarzschild orbits of radius $r_0$ have energy and angular momentum given by \cite{Carroll:2004st,Chandrasekhar:1985kt}
\begin{align}
 E^2=\frac{r_0-2m}{r_0(r_0-3m)},\quad L^2=\frac{r^2_0m}{r_0-3m}. \label{E_0}
\end{align}
The solution to the equations of motion is 
\begin{align}
 r=r_0=\mathrm{constant},\quad \theta(\tau)=\frac{L}{r_0^2}\tau. \label{Sch_circular-polar}
\end{align}
For small $B$, suppose the trajectory of the particle is given by
\begin{align}
 r(\tau)=r_0+Br_1(\tau)+\mathcal{O}\brac{B^2},\quad\theta(\tau)=\frac{L}{r_0^2}\tau+B\theta_1(\tau)+\mathcal{O}\brac{B^2}. \label{ernstE_polar_pert}
\end{align}
Substituting into Eqs.~\Eqref{ernstE_rddot} and \Eqref{ernstE_thetaddot} and solving the equations to linear order in $B$, we obtain\footnote{The procedure is similar to the perturbations considered in \cite{Lim:2014qra}, where the details can be found. Additionally, it is important to note that we have the implicit assumption that $e$ is sufficiently small that terms of the order $\mathcal{O}\brac{eB^2}$ may be neglected.}
\begin{align}
 \ddot{r}_1=&\;-\omega^2r_1+\frac{e(3r_0-8m)}{\sqrt{r_0(r_0-3m)}}\cos\Omega\tau,
\end{align}
where
\begin{align}
 \omega^2=&\;\frac{m(r_0-6m)}{r_0^3(r_0-3m)},\quad\Omega^2=\frac{m^2}{r_0^2\brac{r_0-3m}}. \label{SHO_freq}
\end{align}
Similar to the case of orbits around weakly accelerated black holes \cite{Lim:2014qra}, the radial equation reduces to that of a harmonic oscillator with a periodic driving force. Stable orbits corresponding to $\omega^2>0$ are satisfied by $r_0>6m$, in accordance with the well-known results regarding the stability of Schwarzschild orbits. For the case of charged particles with $e\neq0$, there is an extra term which acts as a periodic driving force.
\par
\begin{figure}[H]
 \begin{center}
  \begin{subfigure}[b]{0.4\textwidth}
   \centering
   \includegraphics{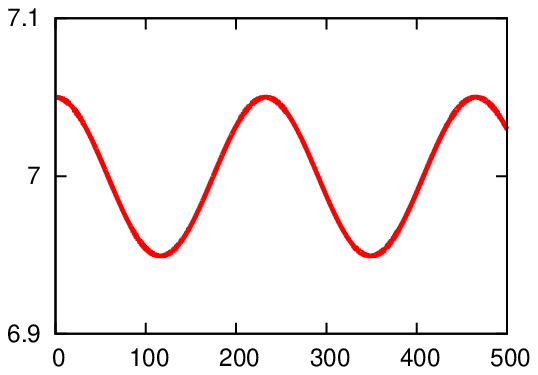}
   \caption{$r$ vs $\tau$ for $e=0$.}
   \label{fig02a}
  \end{subfigure}
  \begin{subfigure}[b]{0.4\textwidth}
   \centering
   \includegraphics{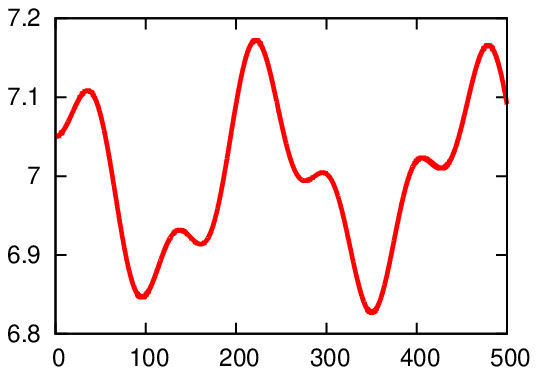}
   \caption{$r$ vs $\tau$ for $e=1$.}
   \label{fig02b}
  \end{subfigure}
   \caption{(Color online) Plot of $r$ vs $\tau$ of perturbations about circular polar orbits of $r_0=7$ (in units where $m=1$) by a weak electric field $B=0.0001$. The plot shows the numerical solutions of $r$ vs $\tau$ for (a) a neutral particle $e=0$, and (b) particle of charge $e=1$. From Eq.~\Eqref{SHO_freq} we can see that the period of oscillation in (a) is $2\pi/\omega\simeq 233$, in agreement with the oscillations seen above. In (b) there is an additional driving force with frequency $\Omega$. The initial conditions are $r(0)=7.05$, $\dot{r}(0)=0$, and $\theta(0)=0$.}
   \label{fig2}
 \end{center}
\end{figure}

 
We can check that the full nonperturbative numerical solutions of \Eqref{ernstE_rddot} and \Eqref{ernstE_thetaddot} are consistent with the above results if they are solved for small $B$. Figure \ref{fig02a} shows the numerical solution of $r$ vs $\tau$ of a neutral particle in nearly circular polar orbit around the black hole. It is possible to check that the particle undergoes simple, undriven oscillation in the radial direction with a period of $2\pi/\omega\simeq 233$, in agreement with Eq.~\Eqref{SHO_freq}.
\par 
Figure \ref{fig02b} shows a similar solution for a charged particle of $e=1$, demonstrating the behavior of a driven oscillator with driving frequency given by $\Omega$.
\par 
Next we we consider the case where the initial circular orbit is in the equatorial plane. In this case the perturbed equations decoupled and are
\begin{align}
 \ddot{r}_1=-\omega^2 r_1,\quad\ddot{\theta}_1=-\Omega^2\theta_1-\frac{e}{\sqrt{r_0(r_0-3m)}}, \label{Ecirc_eq_freq}
\end{align}
where $\omega$ and $\Omega$ are the same as defined in \Eqref{SHO_freq}. This time we see that the equation for $\ddot{\theta}_1$ has a constant term if $e\neq0$. This represents the constant force by the electric field, pushing the particle out of the equatorial plane.

\subsection{Circular orbits in arbitrary field strengths}
For $B$ that is not necessarily small, we can find circular orbits by demanding that $r$ and $\theta$ be constant in Eqs.~\Eqref{ernstE_rddot} and \Eqref{ernstE_thetaddot}. This requires that the energy and angular momentum be given by
\begin{align}
 E=&\;\frac{eBr_0\brac{r_0-2m-m\cos^2\theta_0}}{m\cos\theta_0},\nonumber\\
 \Phi^2=&\;\frac{256e^2B^2r_0^5\sin^4\theta_0\brac{4m+2B^2r_0^3-3mB^2r_0^2-mB^2r_0^2\cos^2\theta_0}}{m^2\cos^2\theta_0\brac{4-B^2r_0^2\sin^2\theta_0}^5},
\end{align}
for some constant $r_0$ and $\theta_0$. 
\par
\begin{figure}
 \begin{center}
  \includegraphics{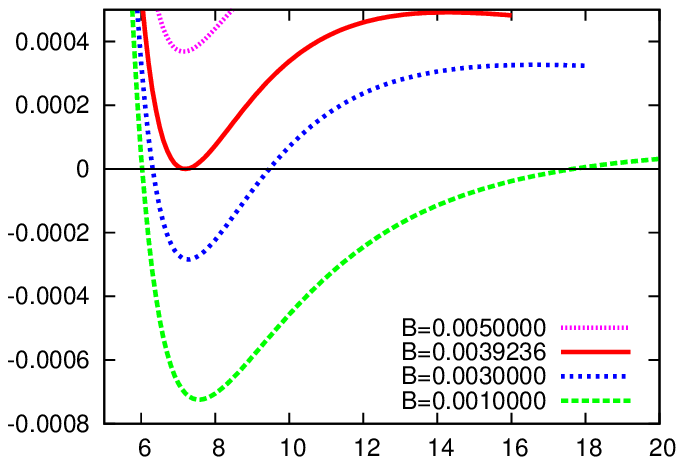}\includegraphics{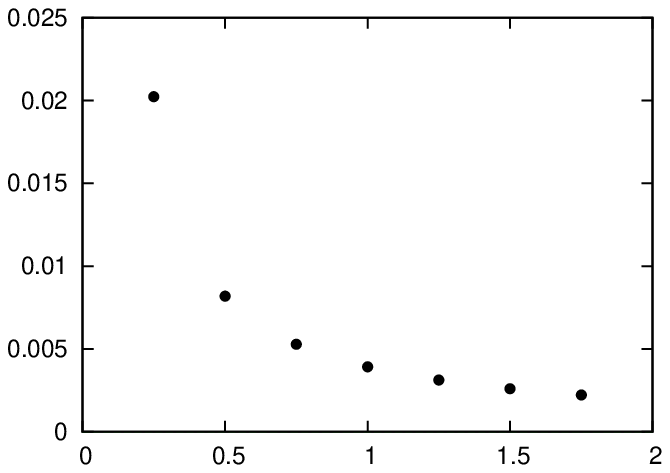}
  \caption{(Color online) \emph{Left}: Plot of $\lambda_+$ vs $r_0$, of various $mB$, with fixed $e=1$ and $m=1$. From the bottom curve, the values of $B$ are 0.001, 0.003, 0.0039236($=B_{\mathrm{crit}}$), and 0.001. \emph{Right}: Plot of $B_{\mathrm{crit}}$ vs $e$.}
  \label{eigens}
 \end{center}
\end{figure}

 
We perturb about the circular orbits by writing
\begin{align}
 r(\tau)=r_0+\varepsilon r_1(\tau)+\mathcal{O}\brac{\varepsilon^2},\quad \theta(\tau)=\theta_0+\varepsilon \theta_1(\tau)+\mathcal{O}\brac{\varepsilon^2}.
\end{align}
Substituting into Eqs.~\Eqref{ernstE_rddot} and \Eqref{ernstE_thetaddot} and expanding to first order gives
\begin{align}
 \frac{\dif^2}{\dif\tau^2}\left(\begin{array}{c}
                                 r_1 \\ \theta_1
                                \end{array}\right)=
                          \left(\begin{array}{cc}
                                 A_{11} & A_{12} \\
                                 A_{21} & A_{22}
                                \end{array}\right)
                          \left(\begin{array}{c}
                                 r_1 \\ \theta_1
                                \end{array}\right),
\end{align}
where $A_{ij}$ are somewhat complicated functions of $r_0$ and $\theta_0$, though it can be handled appropriately with the aid of a symbolic computation software such as MAPLE. We find the normal mode frequencies by solving for the eigenvalues of the above $2\times2$ matrix, which is given by
\begin{align}
 \lambda_\pm=\half\brac{A_{11}+A_{22}\pm\sqrt{(A_{11}-A_{22})^2+4A_{21}A_{12}}}.
\end{align}
The orbit is unstable if one or both of the eigenvalues are positive. Hence, a sufficient condition for instability is that the larger eigenvalue has a range of $r_0$ which is positive. The left-hand plot in Fig.~\ref{eigens} shows $\lambda_+$ as a function of $r_0$ for fixed $e=1$. When the field strength is relatively weak, there is a wide range of $r_0$ where stable orbits are possible. This range shrinks as $mB$ is increased. Beyond the value of $mB_{\mathrm{crit}}\simeq0.00392360$, there are no more stable orbits.
\par 
The particle's charge is another free parameter and hence $B_{\mathrm{crit}}$ should also depend on $e$. The right-hand plot of Fig.~\ref{eigens} shows the dependence of $mB_{\mathrm{crit}}$ on the charge $e$. We can clearly see that the greater the particle's charge, the lower the value of $mB_{\mathrm{crit}}$. Indeed, when the particle's charge is higher, it will experience a greater Coulomb force from the electric field and hence, be more easily rendered unstable.

\section{Charged particles in the magnetic Ernst spacetime} \label{ernstM}

\subsection{Equations of motion}

For the case of the magnetized Ernst spacetime, the vector potential is given by \Eqref{A_M} and the Lagrangian for the charged test particle is
\begin{align}
 \mathcal{L}=&\;\half\sbrac{\Lambda^2\brac{-f\dot{t}^2+\frac{\dot{r}^2}{f}+r^2\dot{\theta}^2}+\frac{r^2\sin^2\theta}{\Lambda^2}\dot{\phi}^2}+eB\frac{r^2\sin^2\theta}{2\Lambda}\dot{\phi}.
\end{align}
In this case, $\partial/\partial t$ and $\partial/\partial\phi$ are still Killing vectors associated with the conserved energy and angular momentum $E$ and $\Phi$. This leads to the first integrals,
\begin{align}
 \dot{t}=&\;\frac{E}{\Lambda^2f},\quad\dot{\phi}=\frac{\Lambda^2}{r^2\sin^2\theta}\brac{\Phi-\frac{eBr^2\sin^2\theta}{2\Lambda}}.\label{ernstM_cons}
\end{align}
The equations of motion for $r$ and $\theta$ are
\begin{align}
 \ddot{r}=&\;\brac{\frac{f'}{2f}-\frac{\partial_r\Lambda}{\Lambda}}\dot{r}^2+f\brac{r+\frac{r^2\partial_r\Lambda}{\Lambda}}\dot{\theta}^2-\frac{2\partial_\theta\Lambda}{\Lambda}\dot{r}\dot{\theta}-\brac{\frac{\partial_r\Lambda}{\Lambda}+\frac{f'}{2f}}\frac{E^2}{\Lambda^4}\nonumber\\
      &\;+\frac{f}{r^3\sin^2\theta}\brac{1-\frac{r\partial_r\Lambda}{\Lambda}}\brac{\Phi-\frac{eBr^2\sin^2\theta}{\Lambda}}^2\nonumber\\
      &\;+\frac{eBf}{r\Lambda}\brac{1-\frac{r\partial_r\Lambda}{2\Lambda}}\brac{\Phi-\frac{eBr^2\sin^2\theta}{\Lambda}},\label{ernstM_rddot}\\
 \ddot{\theta}=&\;\frac{\partial_\theta\Lambda}{\Lambda}\brac{\frac{\dot{r}^2}{r^2f}-\dot{\theta}^2}-2\brac{1+\frac{\partial_r\Lambda}{\Lambda}}\frac{\dot{r}\dot{\theta}}{r}-\frac{E^2\partial_\theta\Lambda}{r^2\Lambda^5f}\nonumber\\
     &\;+\frac{1}{r^4\sin^3\theta}\brac{\cos\theta-\frac{\sin\theta\partial_\theta \Lambda}{\Lambda}}\brac{\Phi-\frac{eBr^2\sin^2\theta}{2\Lambda}}^2\nonumber\\
     &\;+\frac{eB}{r^2\Lambda\sin\theta}\brac{\cos\theta-\frac{\sin\theta\partial_\theta\Lambda}{2\Lambda}}\brac{\Phi-\frac{eBr^2\sin^2\theta}{2\Lambda}}, \label{ernstM_thetaddot}
\end{align}
and the first integral now is
\begin{align}
 -\frac{E^2}{\Lambda^2f}+\Lambda^2\sbrac{\frac{1}{r^2\sin^2\theta}\brac{\Phi-\frac{eBr^2\sin^2\theta}{2\Lambda}}^2+\frac{\dot{r}^2}{f}+r^2\dot{\theta}^2}=\epsilon. \label{ernstM_first}
\end{align}
As expected, setting $B=0$ reduces to the geodesic equations of the Schwarzschild spacetime, while $m=0$ gives the equations for the Melvin magnetic universe. Setting $\Lambda=1$ describes charged particles in the test field regime considered in \cite{Frolov:2010mi}, where the magnetic field is sufficiently weak so as not to influence the spacetime curvature but only the charged particle through Lorentz interaction. A common special case of interest is equatorial motion where $\theta=\pi/2$ and $\dot{\theta}=0$. Then we have $\ddot{\theta}=0$ and the motion is confined to the equatorial plane. Thus, the $r$ equation can be solved by direct integration of Eq.~\Eqref{ernstM_first}. Furthermore, due to the symmetry given in \Eqref{ernstM_sym},  by a reasoning analogous to the electric Ernst case, we shall assume $B>0$ and $e>0$ without loss of generality.
\par 
Equation.~\Eqref{ernstM_first} can be expressed as an effective potential formulation in the following form,
\begin{align}
 \Lambda^4\brac{\dot{r}^2+r^2f\dot{\theta}^2}=E^2-V^2_{\mathrm{eff}},\quad V^2_{\mathrm{eff}}=\Lambda^4f\sbrac{\frac{1}{r^2\sin^2\theta}\brac{\Phi-\frac{eBr^2\sin^2\theta}{2\Lambda}}^2-\frac{\epsilon}{\Lambda^2}}. \label{ernstM_Veff}
\end{align}

\subsection{Circular orbits in the test field approximation}

In the context of astrophysics, it is typically sufficient to consider magnetic fields which are sufficiently weak so that they do not influence the spacetime curvature \cite{Aliev:1989}. In our present paper we shall call this case the ``test field regime'', where, lacking the influence from the magnetic field, the spacetime metric is simply the Schwarzschild metric [$\Lambda\rightarrow1$ in Eq.~\Eqref{Ernst_metric}], but the vector potential associated with the magnetic field is still given by \Eqref{A_M}.
\par 
Previous works such as \cite{Aliev:2002nw,Perti:2004,Frolov:2010mi} considered the motion of charged particles in the test field regime. The only influence of the magnetic field on the charged particle is via the Lorentz interaction. This regime is contained within our present equations of motion. The resulting equations of motion that follow from the test field regime can be easily obtained by setting $\Lambda\rightarrow 1$ in Eqs.~\Eqref{ernstM_cons}--\Eqref{ernstM_first}. However we wish to have a more rigorous description by seeking an appropriate parameter that will allow a more explicit transition from the test field regime from the full Ernst metric.
\par 
We begin by making the following observation: based on the order-of-magnitude estimations of \cite{Frolov:2010mi}, even in the test field regime a charged particle may still experience a significant Lorentz force that influences its orbit. In other words, $B$ is sufficiently small such that it does not produce spacetime curvature ($\Lambda\simeq 1$), but the Lorentz interaction is still present; hence, $eB$ is non-negligible. With these considerations, we write
\begin{align}
 B=\frac{b}{e},
\end{align}
and expand Eqs.~\Eqref{ernstM_cons}, \Eqref{ernstM_rddot}, \Eqref{ernstM_thetaddot} and \Eqref{ernstM_first} in powers of $1/e$. At zeroth order, we have $B\rightarrow 0$, leading to $\Lambda\rightarrow 1$, but the Lorentz interaction $eB=b$ is still present. In this order we recover the test field regime considered in \cite{Frolov:2010mi}. Terms of the order $\mathcal{O}(1/e)$ and higher come in as the gravitational correction due to the spacetime curvature induced by the magnetic field.
\par
\begin{figure}
 \begin{center}
  \includegraphics{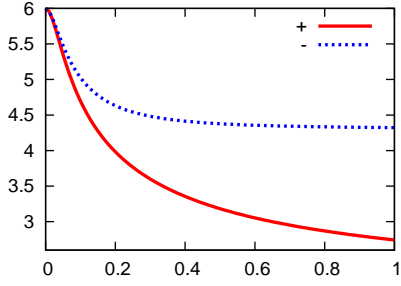}
  \caption{(Color online) Plots of ISCO radii vs $\Phi=\Phi_\pm$, in units where $m=1$. The lower (solid) curve represents the ISCO radius for $\Phi=\Phi_+$, while the upper (dotted) curve corresponds to $\Phi=\Phi_-$.}
  \label{fig04iscos_test}
 \end{center}
\end{figure}

 
As an example, we consider circular equatorial orbits in the magnetic Ernst spacetime. Taking $r=r_0=\mathrm{constant}$ and $\theta=\pi/2$ in the full equations of motion \Eqref{ernstM_rddot}--\Eqref{ernstM_first}, we find that the energy corresponding to a circular orbit is
\begin{align}
 E^2=&\;\frac{(r_0-2m)^2\brac{4e^2+r_0^2b^2}^3\brac{4\Phi e^2+\Phi r_0^2b^2-2e^2br_0^2}\brac{\Phi r_0^2b^2-4\Phi e^2-2e^2br_0^2}}{256e^8r_0^3\brac{3r_0^2mb^2-4me^2-2r_0^3b^2}}.\label{ernstM_Ecirc}
\end{align}
The corresponding values of $\Phi$ may be found by substituting \Eqref{ernstM_Ecirc} into \Eqref{ernstM_first} and solving the quadratic equation for $\Phi$, for which there are two solutions
\begin{align}
  \Phi_\pm=&\;\frac{2e^2r_0\sbrac{r_0(3r_0-5m)b^3+4e^2r_0mb\pm2\sqrt{K}}}{(4e^2+r_0^2b^2)\sbrac{r_0^2(3r_0-5m)b^2+4e^2(r_0-3m)}}. \label{ernstM_Phicirc}
\end{align}
where
\begin{align}
 K=&\;4r_0^2e^2\sbrac{(r_0-2m)^2e^2+2r_0^2-12r_0m+14m^2}b^2-r_0^4(3r_0-5m)(2r_0-3m)b^4\nonumber\\
   &+16e^4m(r_0-3m).
\end{align}
For concreteness, we analyze the innermost stable circular orbits (ISCOs) in the test field regime. These are orbits which are marginally stable, where all circular orbits with radius less than the ISCO are unstable and those with radii larger than it are stable. The stability of the circular orbits is determined by perturbing about the circular orbit solution by writing
\begin{align}
 r(\tau)=r_0+\varepsilon r_1(\tau).
\end{align}
With this ansatz, perturbing Eq.~\Eqref{ernstM_rddot} to first order in $\varepsilon$ reduces to
\begin{align}
 \ddot{r}_1=&-\lambda r_1,\quad\lambda=k_1E^2+k_2\Phi^2+k_3\Phi+k_4,
\end{align}
where 
\begin{align}
 k_1=&\;\frac{512\sbrac{\brac{31mr_0-9r_0^2-27m^2}r_0^4b^2+4\brac{r_0^2+14m^2-10mr_0}e^2r_0^2b^2-16me^4(r_0-m)}}{r_0^2(4e^2+r_0^2b^2)^6(r_0-2m)^2},\nonumber\\
 k_2=&\;\frac{16e^2r_0^2b^2(r_0-2m)-r_0^4(3r_0-8m)b^4+16e^4(3r_0-8m)}{(4e^2+r_0^2b^2)^2},\nonumber\\
 k_3=&\;\frac{4b^3\sbrac{(3r_0^2-8r_0m)b^2+8e^2}e^2}{(4e^2+r_0^2b^2)^3},\nonumber\\
 k_4=&\;\frac{4b^2\sbrac{r_0b^2(8m-3r_0)+4e^2}}{(4e^2+r_0^2b^2)^3},
\end{align}
and $E$ and $\Phi$ are given by \Eqref{ernstM_Ecirc} and \Eqref{ernstM_Phicirc}. As usual, stable circular orbits have $\lambda>0$ and unstable ones have $\lambda<0$. The expression for $\lambda$ is expanded in powers of $1/e$ to give
\begin{align}
 \lambda_\pm\simeq\lambda_\pm^{(0)}+\frac{\lambda_\pm^{(2)}}{e^2}+\mathcal{O}\brac{1/e^4}, \label{lambda_exp}
\end{align}
where the $\pm$ signs correspond to the two distinct choices of $\Phi=\Phi_\pm$. To find an ISCO, we solve $\lambda_\pm=0$, where the upper and lower signs lead to two distinct ISCO radii $r_0=r_\pm$. At zeroth order in $1/e$, which is the test field regime, the results of \cite{Frolov:2010mi} are reproduced, shown in Fig.~\ref{fig04iscos_test}. Particularly for $b\neq 0$, we typically have $r_\pm<6m$, i.e., the typical ISCO radius for  nonzero magnetic fields is smaller than the Schwarzschild ISCO radius.
\par
\begin{figure}
 \begin{center}
   \begin{subfigure}[b]{0.4\textwidth}
   \centering
   \includegraphics{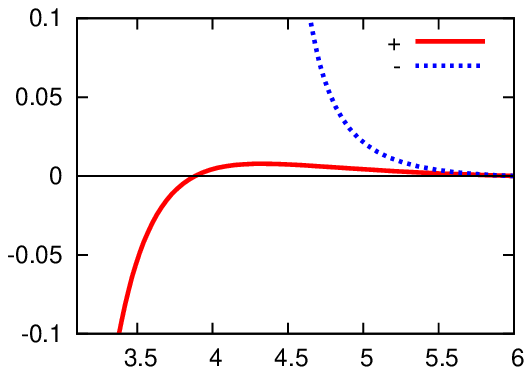}
   \caption{$\lambda_\pm^{(2)}$ vs $r_\pm$.}
   \label{iscos_grav_r}
  \end{subfigure}
  \begin{subfigure}[b]{0.4\textwidth}
   \centering
   \includegraphics{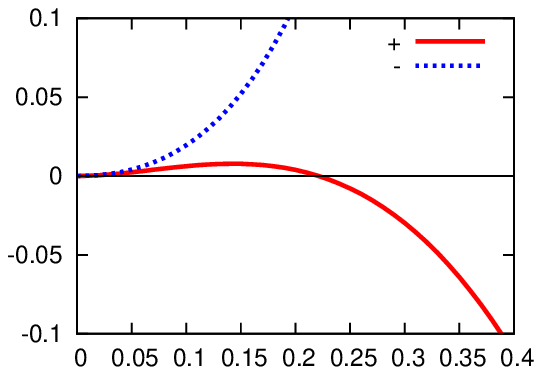}
   \caption{$\lambda_\pm^{(2)}$ vs $b$.}
   \label{iscos_grav_b}
  \end{subfigure}
  \caption{(Color online) Plots of $\lambda_\pm^{(2)}$ against $r$ and $b$, plotted in units where $m=1$. The solid curves represent $\lambda_+$ and $r_+$ while the dotted curves correspond to $\lambda_-$ and $r_-$.}
  \label{iscos_grav}
 \end{center}
\end{figure}

 
Next we introduce the gravitational corrections to the above results. For ease of exposition, we shall denote the ISCO radius evaluated in the test field regime as the ``test field ISCO radius'', and the ISCO radius when taking first-order corrections into account as ``gravitationally corrected ISCO radius''. For an ISCO with given $r_\pm$ and $b$, we find the gravitational correction to $\lambda$ by calculating the second term in \Eqref{lambda_exp}. For the background ISCO parameters, by their definition of being marginally stable orbits, we have $\lambda_\pm^{(0)}=0$. Thus, if $\lambda_\pm^{(2)}$ is found to be positive, the ISCO orbits are stabilized by gravitational corrections. Conversely, if we find $\lambda_\pm^{(2)}$ to be negative, then the ISCO orbits are rendered unstable. The particular expression of $\lambda_\pm^{(2)}$ is again too complicated to be displayed here, but the dependence of $b$ or, equivalently, $r_\pm$ is shown in Fig.~\ref{iscos_grav}.
\par 
We note that in the case $\Phi_-$, we have $\lambda_-^{(2)}>0$; hence, circular orbits in the test field ISCO radius are stabilised by gravitational corrections. In other words, test field ISCO radius $r_-$ is now stable and the gravitationally corrected ISCO radius is slightly smaller than the test field ISCO radius.
\par 
The case $\Phi_+$ is more interesting. As seen in Fig.~\ref{iscos_grav_b}, for relatively small $b$, we still have $\lambda_+^{(2)}>0$; the gravitationally corrected ISCOs are also further inwards. However, beyond a critical value of approximately $b\simeq0.22134$, the value of $\lambda_+^{(2)}$ is negative. Hence, in this case gravitational correction actually destabilizes the test field ISCOs. When gravity is taken into account, the gravitationally corrected ISCO radius is further away from the black hole.

\subsection{Curly orbits: Trochoidlike trajectories in the Melvin spacetime}

In Ref.~\cite{Frolov:2010mi}, Frolov and Shoom considered an interesting behavior where the trajectory of charged timelike particle curls up into a cycloidlike (or more generally, trochoidlike) motion. This behavior is due to the fact that, for motion in the equatorial plane, the charged particle experiences a gravitational force in the direction orthogonal to the magnetic field lines. As explained in the previous subsection, they have considered the test field ($\Lambda\rightarrow 1$) regime, where only the Schwarzschild black hole contributes to the orthogonal gravitational force, and the magnetic field is solely responsible for the Lorentz force.
\par 
Here we consider a similar effect for the Melvin magnetic universe.\footnote{We expect that the presence of a black hole does not add any notable physical difference to bound trochoidlike orbits, aside from a stronger gravitational force towards the center. Hence we avoid unnecessarily cumbersome equations by taking $m=0$.} While there is no black hole present in this case, the magnetic field itself exerts a gravitational force on the particle, in addition to providing the Lorentz interaction. The curling-up behavior is characterized by the fact that $\dot{\phi}$ changes sign in Eq.~\Eqref{ernstM_cons}. This sign change occurs when $r=r_*$, where
\begin{align}
 r_*=\frac{1}{\sin\theta}\sqrt{\frac{4\Phi}{2eB-\Phi B^2}}.
\end{align}
In the following we will consider only bounded motion in which the particle is confined within the range $r_{\mathrm{min}}\leq r\leq r_{\mathrm{max}}$, where the boundaries are defined by $E^2=V^2_{\mathrm{eff}}$ in Eq.~\Eqref{ernstM_Veff}. If we further consider motion confined in the equatorial plane, it is possible derive approximate solutions representing the cycloid- or trochoidlike motion if we consider perturbations about circular orbits in the Melvin spacetime. Hence, we solve Eq.~\Eqref{ernstM_rddot} by substituting $\theta=\pi/2$ and taking
\begin{align}
 r(\tau)=r_0+\varepsilon r_1(\tau)+\mathcal{O}\brac{\varepsilon^2}. \label{Melvin_perturb}
\end{align}
At zeroth order, demanding that $r_0$ is constant gives
\begin{align}
 E^2=\frac{\brac{16\Phi^2-4e^2B^2r_0^4+4eB^3\Phi r_0^4-B^4\Phi^2r_0^4}\brac{4+B^2r_0^2}}{512r_0^4B^2}.
\end{align}
At first order, the equations of motion reduce to $\ddot{r}_1=-\omega^2r_1$, where
\begin{align}
 \omega^2=\frac{2\brac{3\Phi^2B^6r_0^6-12eB^5\Phi r_0^6+12e^2B^6r_0^6-16B^2\Phi^2r_0^2+128\Phi^2}}{(4+B^2r^2)^3r_0^4}.
\end{align}
Substituting \Eqref{Melvin_perturb} into the $\dot{\phi}$ equation, we find
\begin{align}
 \dot{\phi}=\alpha_0+\alpha_1\varepsilon\cos\omega\tau+\mathcal{O}\brac{\varepsilon^2},
\end{align}
where we have used the solution $r_1=\cos\omega\tau$, and
\begin{align}
 \alpha_0=\frac{\brac{4\Phi+B^2\Phi r_0^2-2eBr_0^2}\brac{4+B^2r_0^2}}{16r_0^2},\quad\alpha_1=\frac{B^4\Phi r_0^4-2eB^3r_0^4-16\Phi}{8r_0^3}.
\end{align}
Therefore, the approximate solutions to the equations of motion are, to first order in $\varepsilon$,
\begin{align}
 r(\tau)=&\;r_0+\varepsilon\cos\omega\tau,\nonumber\\
 \phi(\tau)=&\;\alpha_0\tau+\frac{\alpha_1\varepsilon}{\omega}\sin\omega\tau.
\end{align}
This describes the locus of a trochoid, where there are three possible types depending on the relationship among its parameters. Let us define
\begin{align}
 \eta=-\frac{\alpha_0}{\alpha_1\varepsilon}.
\end{align}
If $\eta=1$, the motion is that of a usual cycloid, where the trajectory forms sharp cusps at maximum $r$. In more general cases, $\eta<1$ is known as the prolate cycloid, and $\eta>1$ corresponds to a curtate cycloid.
\par
\begin{figure}
  \begin{center}
  \begin{subfigure}[b]{1\textwidth}
   \centering
   \includegraphics[scale=1]{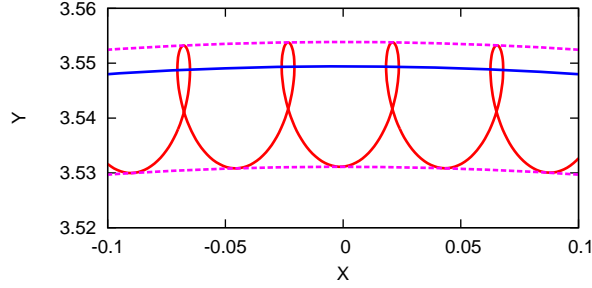}
   \caption{$E=1.0079$, $\eta\simeq0.608<1$.}
   \label{curly_thr}
  \end{subfigure}
  \begin{subfigure}[b]{1\textwidth}
   \centering
   \includegraphics[scale=1]{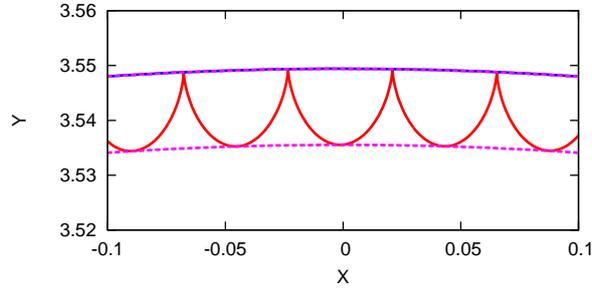}
   \caption{$E=1.007874016$, $\eta\simeq0.997\sim 1$.}
   \label{curly_crit}
  \end{subfigure}
  \begin{subfigure}[b]{1\textwidth}
   \centering
   \includegraphics[scale=1]{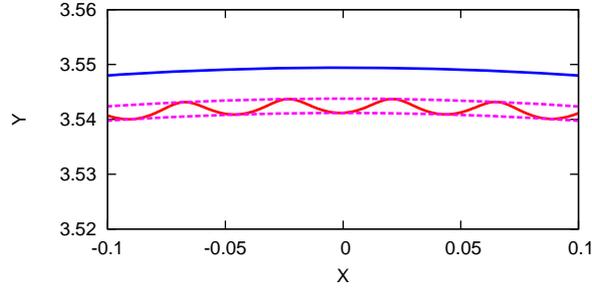}
   \caption{$E=1.00785912$, $\eta\simeq5.329>1$.}
   \label{curly_wav}
  \end{subfigure}
  \caption{(Color online) Numerical solutions for $B=0.05$, $\Phi=5$, $e=16$. The dashed lines are the boundary of the allowed motion defined by $E^2=V^2_{\mathrm{eff}}$ while the dotted curve represents $r=r_*$. In (a) we show the case corresponding to a prolate cycloid with $\eta<1$, (b) is the common cycloid, and (c) is the curtate cycloid. As we see in (b) the curve $r=r_*$ coincides with the maximum allowed $r$ of the motion, producing the cycloidlike trajectory. For each solution, for a given set of parameters $(E,\Phi,B,e)$, the value $r_0$ is calculated from $\dif\brac{V^2_{\mathrm{eff}}}/\dif r=0$ and $\varepsilon$ may be calculated from $E^2=V^2_{\mathrm{eff}}$.}
  \label{curlies}
  \end{center}
\end{figure}

 
We can verify the above solutions with the numerical solutions of the fully non-perturbative equations with the appropriate range of parameters. Fig.~\ref{curlies} shows a solution for $B=0.05$, $\Phi=5$ and $e=16$. It appears that, nearly circular orbits with $r=r_*$ lying close to the vicinity of the oscillation typically occur for highly charged particles; hence, the choice $e=16$ in Fig.~\ref{curlies}.  
\par
\begin{figure}
 \begin{center}
   \begin{subfigure}[b]{0.4\textwidth}
   \centering
   \includegraphics{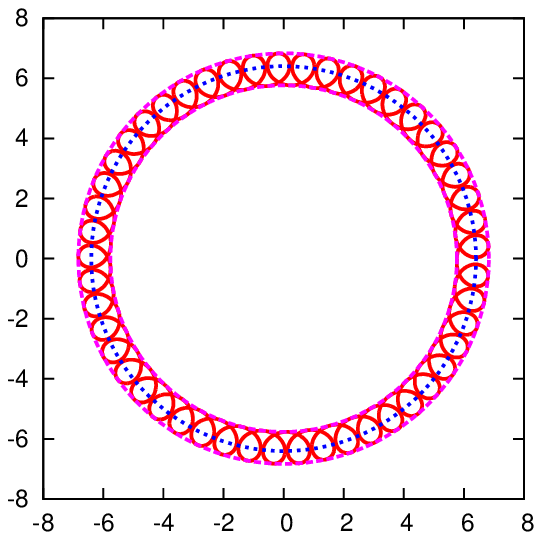}
   \caption{$(B,e,E,\Phi)=\brac{0.05,5,\sqrt{1.07},5}$}
   \label{Mcurl_example1}
  \end{subfigure}
  \begin{subfigure}[b]{0.4\textwidth}
   \centering
   \includegraphics{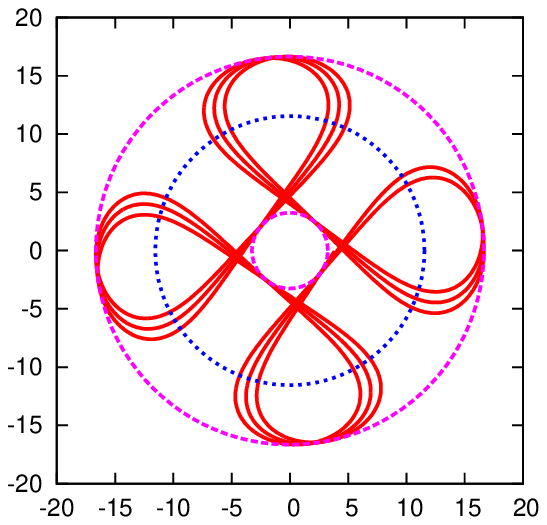}
   \caption{$(B,e,E,\Phi)=\brac{0.1,1,1.78,5}$}
   \label{Mcurl_example2}
  \end{subfigure}
  \begin{subfigure}[b]{0.4\textwidth}
   \centering
   \includegraphics{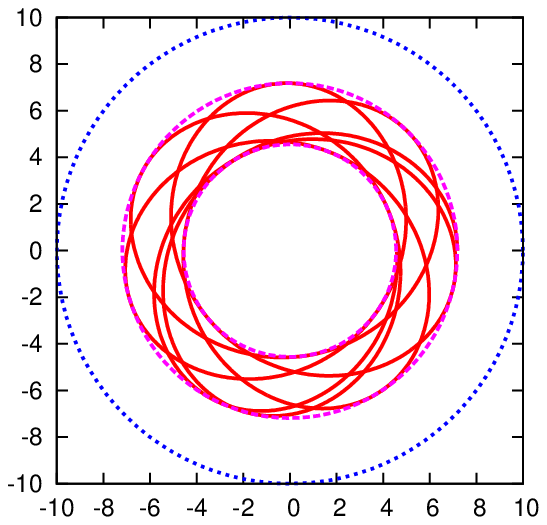}
   \caption{$(B,e,E,\Phi)=\brac{0.2,1,1.6,5}$}
   \label{Mcurl_example3}
  \end{subfigure}
  \begin{subfigure}[b]{0.4\textwidth}
   \centering
   \includegraphics{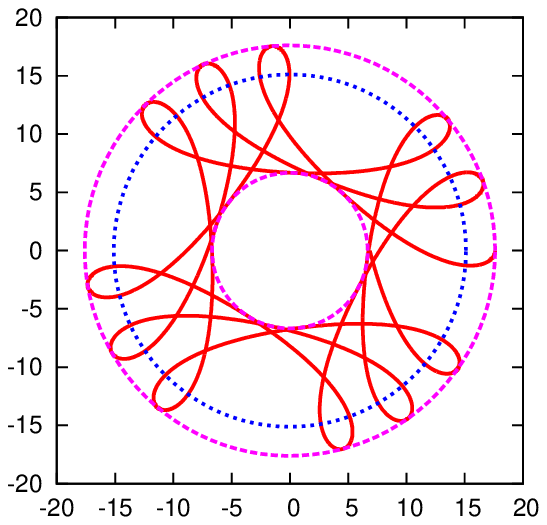}
   \caption{$(B,e,E,\Phi)=\brac{0.05,1,1.2,5}$}
   \label{Mcurl_example4}
  \end{subfigure}
  \caption{(Color online) Trochoidlike trajectories for various orbital parameters. The dashed circles indicate the boundary of the allowed motion defined by $E^2=V^2_{\mathrm{eff}}$ while the dotted circle represents $r=r_*$, the turning point of $\phi$. In (c) it can be seen that $r_*$ lies beyond the accessible range and, therefore, does not curl up.}
  \label{Mcurl_example}
 \end{center}
\end{figure}

 
Of course, the above calculations are valid for nearly circular orbits where the radius is close to $r_0$. Moving beyond the perturbative range, we can obtain the curly orbits from numerical solutions. Some examples are shown in Fig.~\ref{Mcurl_example}.
\par
For the rest of this section we briefly consider the critical, cycloidlike orbits in the nonperturbative range. This class can be easily studied since they can be characterized by simple equations. It is clear that the trajectories develop cycloidlike cusps if $r_*$ coincides with $r_{\mathrm{max}}$. Since by definition $V^2_{\mathrm{eff}}\brac{r=r_{\mathrm{max}}}=E^2$, it follows that
\begin{align}
 r_*=\frac{2\sqrt{E-1}}{B},\quad\Phi=\frac{2e(E-1)}{BE}. \label{cusp_relations}
\end{align}
We conclude that such orbits exist under the condition $E>1$, and $\Phi>0$. (Recall that due to Eq.~\Eqref{ernstM_sym}, we may assume $B>0$ and $e>0$ without loss of generality.) Following the spirit of \cite{Levin:2008mq}, one may further understand the structure of the orbits by identifying the periodic orbits since any generic orbit appears like perturbations of periodic ones. Periodic orbits are defined as trajectories which return precisely to their initial conditions after a finite proper time; thus, the orbits periodically retrace the same path repeatedly.
\par
\begin{figure}
 \begin{center}
   \begin{subfigure}[b]{0.4\textwidth}
   \centering
   \includegraphics{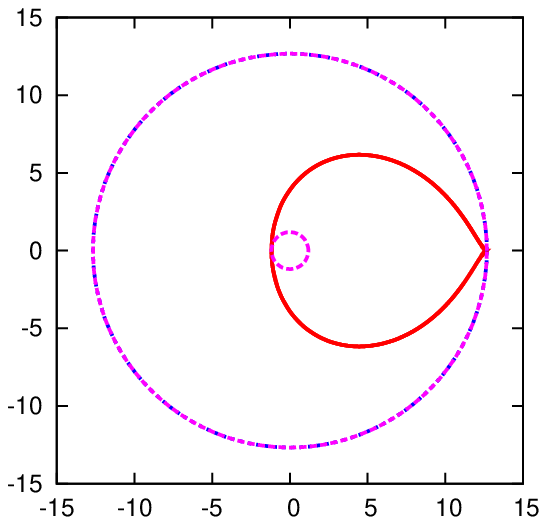}
   \caption{$n=1,\quad E=4.6151$}
   \label{curly_n1}
  \end{subfigure}
  \begin{subfigure}[b]{0.4\textwidth}
   \centering
   \includegraphics{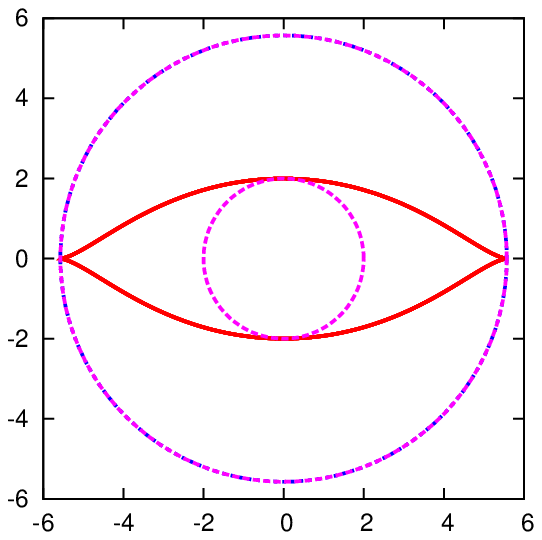}
   \caption{$n=2,\quad E=1.69857$}
   \label{curly_n2}
  \end{subfigure}
  \begin{subfigure}[b]{0.4\textwidth}
   \centering
   \includegraphics{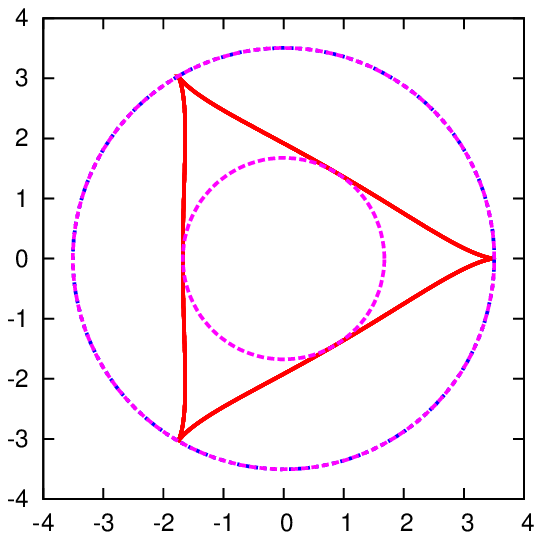}
   \caption{$n=3,\quad E=1.2767$}
   \label{curly_n3}
  \end{subfigure}
  \begin{subfigure}[b]{0.4\textwidth}
   \centering
   \includegraphics{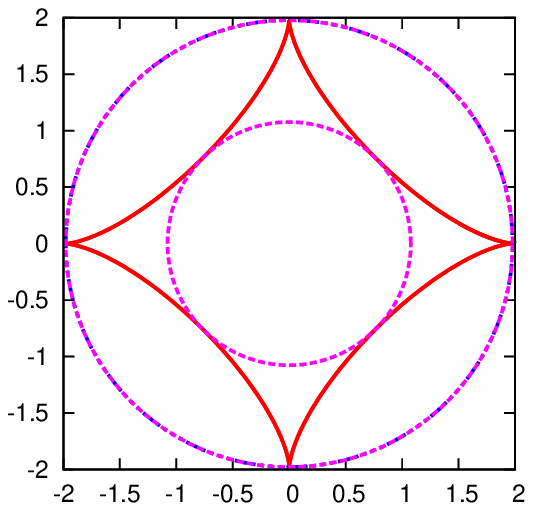}
   \caption{$n=4,\quad E=1.0882$}
   \label{curly_n4}
  \end{subfigure}
  \caption{(Color online) Periodic cuspy orbits for $B=0.3$, and $e=1$.}
  \label{curly_periodics}
 \end{center}
\end{figure}

 
Since, for equatorial orbits, we have $\theta=\pi/2$ and $\dot{\theta}=0$, we can take Eq.~\Eqref{ernstM_Veff} as the first integral for the $r$ motion. Together with the equation for $\dot{\phi}$ from Eq.~\Eqref{ernstM_cons}, we can obtain a trajectory analytically by integrating
\begin{align}
 \frac{\dot{r}}{\dot{\phi}}=\frac{\dif r}{\dif\phi}=\frac{r^2\sqrt{\frac{E^2}{\Lambda^4}-\frac{1}{r^2}\brac{\Phi-\frac{eBr^2}{2\Lambda}}^2+\frac{\epsilon}{\Lambda^2}}}{\Lambda^2\brac{\Phi-\frac{eBr^2}{2\Lambda}}}. \label{cuspy_eom}
\end{align}
We can classify the periodic orbits by the number of cusps $n$, formed before the particle returns to the initial conditions. A particle moves between cusps by starting from $r_{\mathrm{max}}$, reaching a turning point at $r_{\mathrm{min}}$, then returning to $r_{\mathrm{max}}$ again. The orbit will be periodic if the difference in $\phi$ between the cusps is some rational fraction of $2\pi$. In terms of \Eqref{cuspy_eom},
\begin{align}
 2\int_{r_{\mathrm{min}}}^{r_{\mathrm{max}}}\frac{\Lambda^2\brac{\Phi-\frac{eBr^2}{2\Lambda}}\dif r}{r^2\sqrt{\frac{E^2}{\Lambda^4}-\frac{1}{r^2}\brac{\Phi-\frac{eBr^2}{2\Lambda}}^2+\frac{\epsilon}{\Lambda^2}}}=\Delta\phi=\frac{2\pi}{n}, \label{cuspy_soln}
\end{align}
for some integer $n$. In practice, one can find a periodic orbit in the following way: Given parameters $(B,e,E)$, the angular momentum $\Phi$ and $r_*=r_{\mathrm{max}}$ are determined from \Eqref{cusp_relations}, while $r_{\mathrm{min}}$ is determined from solving $E^2=V^2_{\mathrm{eff}}$. With these parameters one can find $\Delta\phi$ from Eq.~\Eqref{cuspy_soln}. Periodic orbits are found by tuning one of the parameters, say, $E$ for some fixed $B$ and $e$ to find an integer solution for which $\Delta\phi=2\pi/n$. In Fig.~\ref{curly_periodics}, we show an example of periodic orbits for $n$ ranging from 1 to 4. It appears that for fixed $B$ and $e$, the number of cusps increase as the energy $E$ decreases.

\section{Neutral particles} \label{neutral}

\subsection{Effective potential for neutral particles}
For the case of neutral particles, the trajectories are then purely determined by the spacetime curvature, and they are not affected by the electromagnetic field. Hence for $e=0$, we see that the geodesic equations for electric and magnetic Ernst spacetime becomes identical. In particular, Eq.~\Eqref{ernstE_first}, now identical to \Eqref{ernstM_first} can be cast into an effective potential form
\begin{align}
 \Lambda^4\brac{\dot{r}^2+r^2f\dot{\theta}^2}=&E^2-V^2_{\mathrm{eff}},\quad V^2_{\mathrm{eff}}=\Lambda^4f\brac{\frac{\Phi^2}{r^2\sin^2\theta}-\frac{\epsilon}{\Lambda^2}}. \label{ernstN_first}
\end{align}
To see the structure of the effective potential for photon orbits, we set $\epsilon=0$ in Eq.~\Eqref{ernstN_first} and find curves in the $(r,\theta)$ plane which satisfies $E^2=V^2_{\mathrm{eff}}$. These curves correspond to points where $\dot{r}=\dot{\theta}=0$ and serve as the boundary of the region accessible by the photon.
\par
\begin{figure}
 \begin{center}
  \includegraphics[scale=0.34]{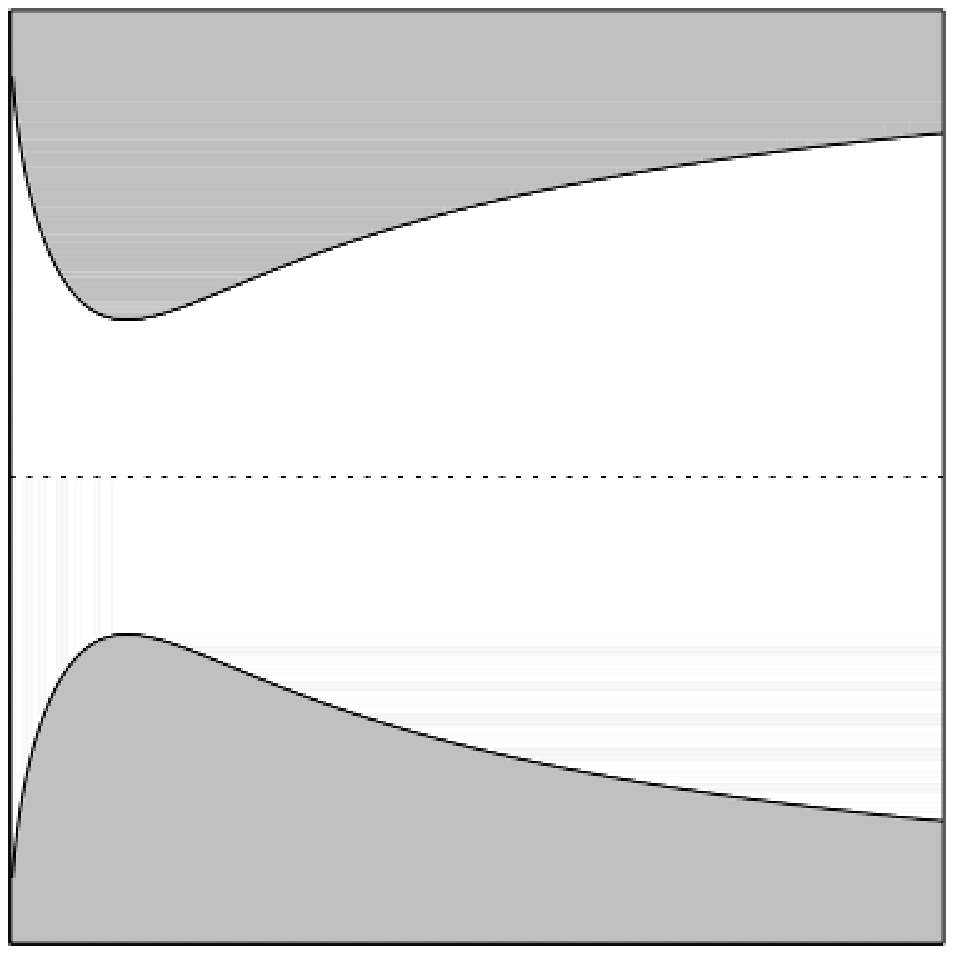}
  \includegraphics[scale=0.34]{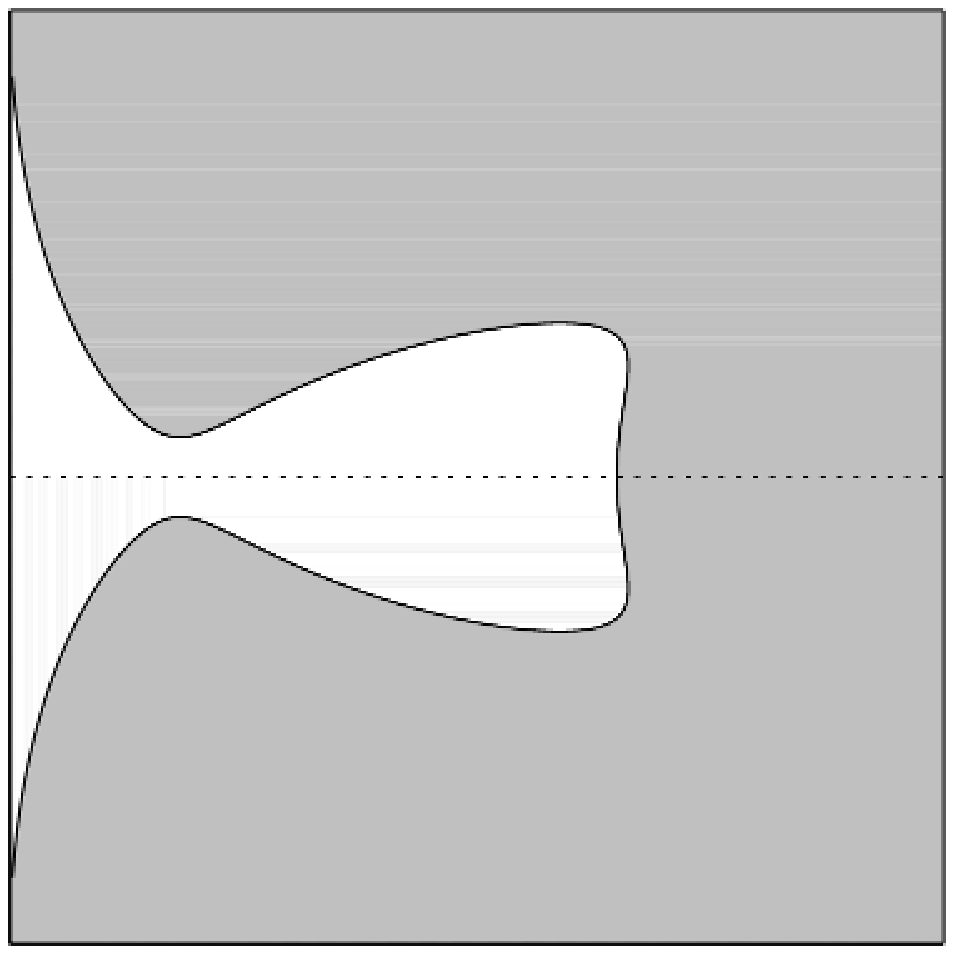}
  \includegraphics[scale=0.34]{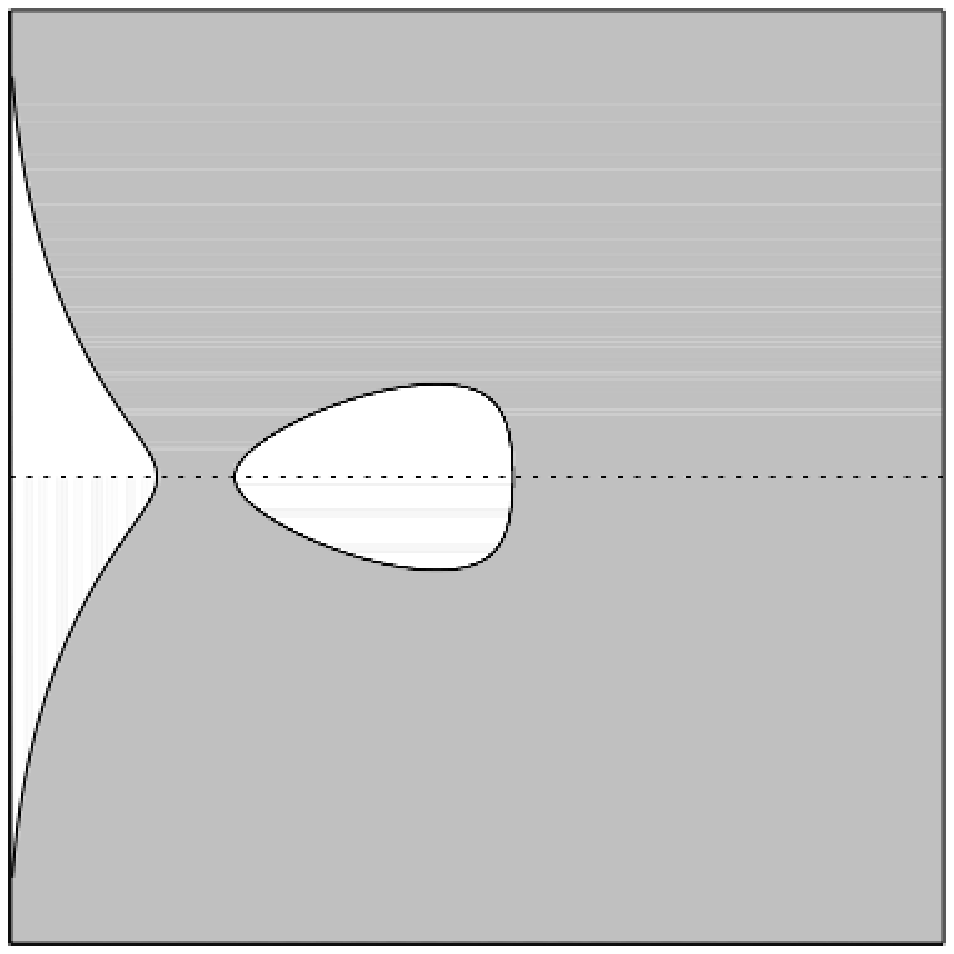}
  \includegraphics[scale=0.34]{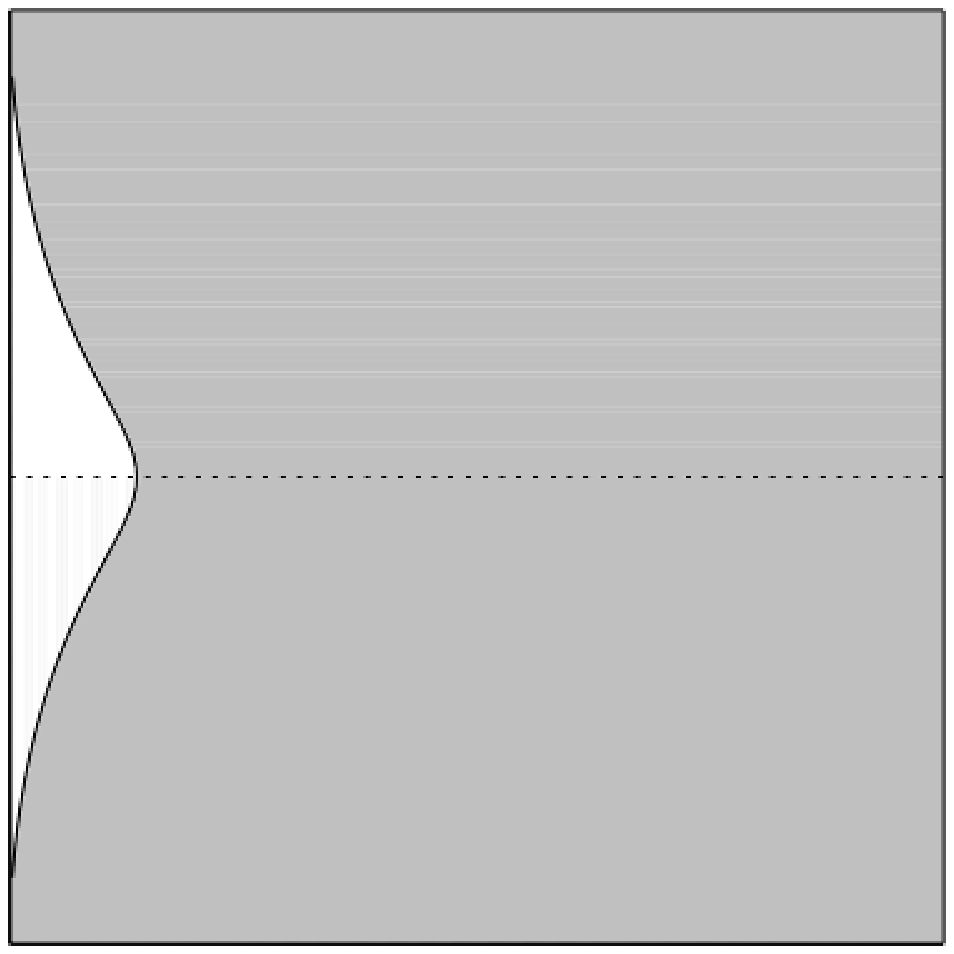}
  \caption{Effective potential for photon orbits for $m=1$, $E=0.669$ and $\Phi=3$. From left to right, the field strength is $B=0, 0.17, 0.175$ and $0.18$. The shaded regions indicate areas where $V^2_{\mathrm{eff}}>E^2$ and, hence, not accessible to the particles. Horizontal and vertical axes  respectively, correspond to $r$ and $\theta$. The horizontal dotted line through the center indicates the equator at $\theta=\pi/2$. The ranges for the above plots are $r\in[2,10]$ and $\theta\in[0,\pi]$.}
  \label{NV_eff_Ph}
 \end{center}
\end{figure}

 
Figure \ref{NV_eff_Ph} shows a typical structure of the effective potential of photon orbits in the Ernst spacetime, where the shaded region indicates areas not accessible to the photon of a given $E$ and $\Phi$. The first diagram from the left is the plot for $B=0$, which is simply the well-known effective potential for Schwarzschild orbits. The following diagrams towards the right show the effect of increasing $B$ for fixed $E$ and $\Phi$, where we see the ``neck'' gradually pinches off as a potential barrier forms, creating an isolated finite potential well in the third diagram. The potential well vanishes if $B$ is increased further. The critical value of $B$ where the potential value shrinks to a single point corresponds to the case of circular photon orbits of constant $r$ and $\theta$, which we will study in detail in the following subsection.
\par
\begin{figure}
 \begin{center}
  \includegraphics[scale=0.34]{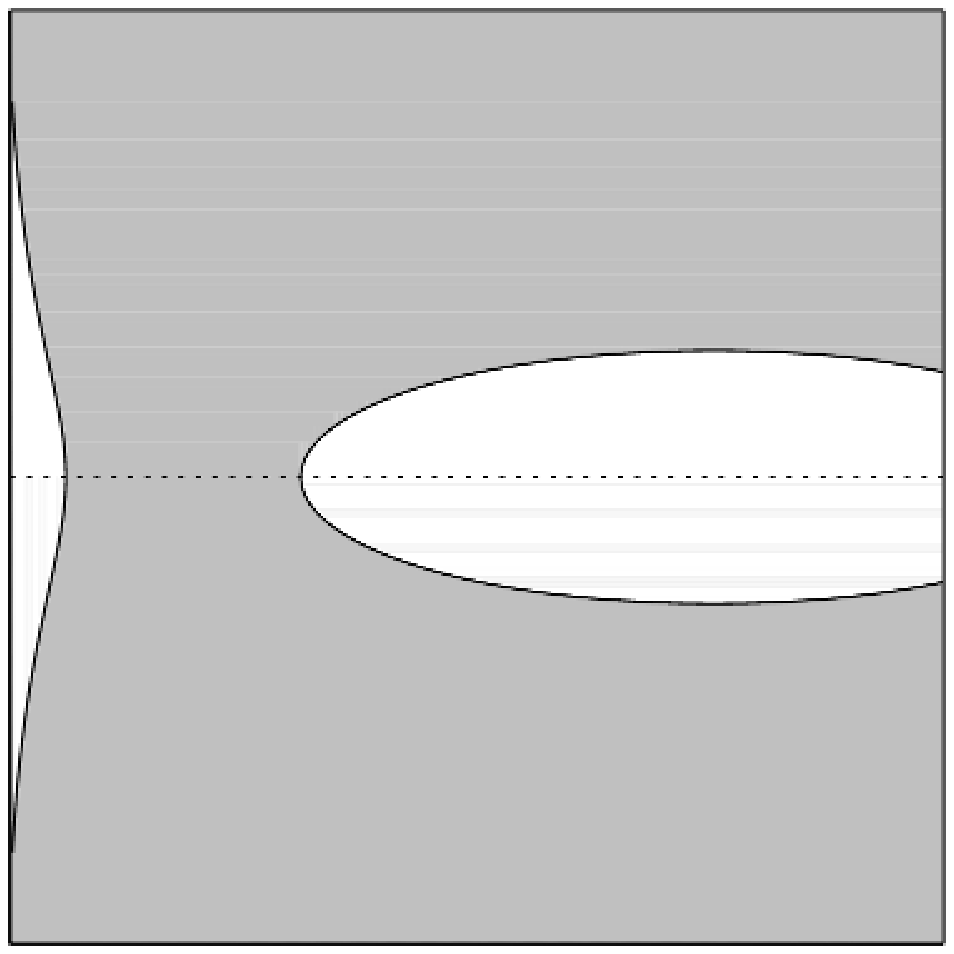}
  \includegraphics[scale=0.34]{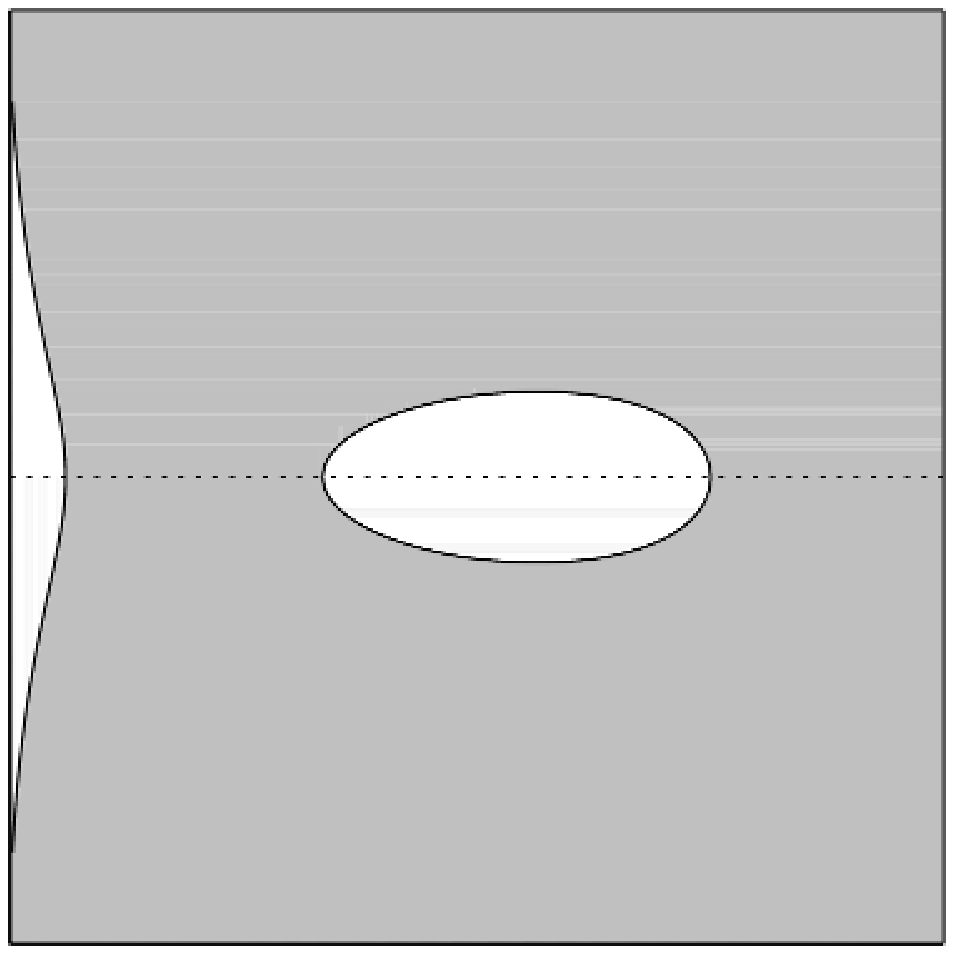}
  \includegraphics[scale=0.34]{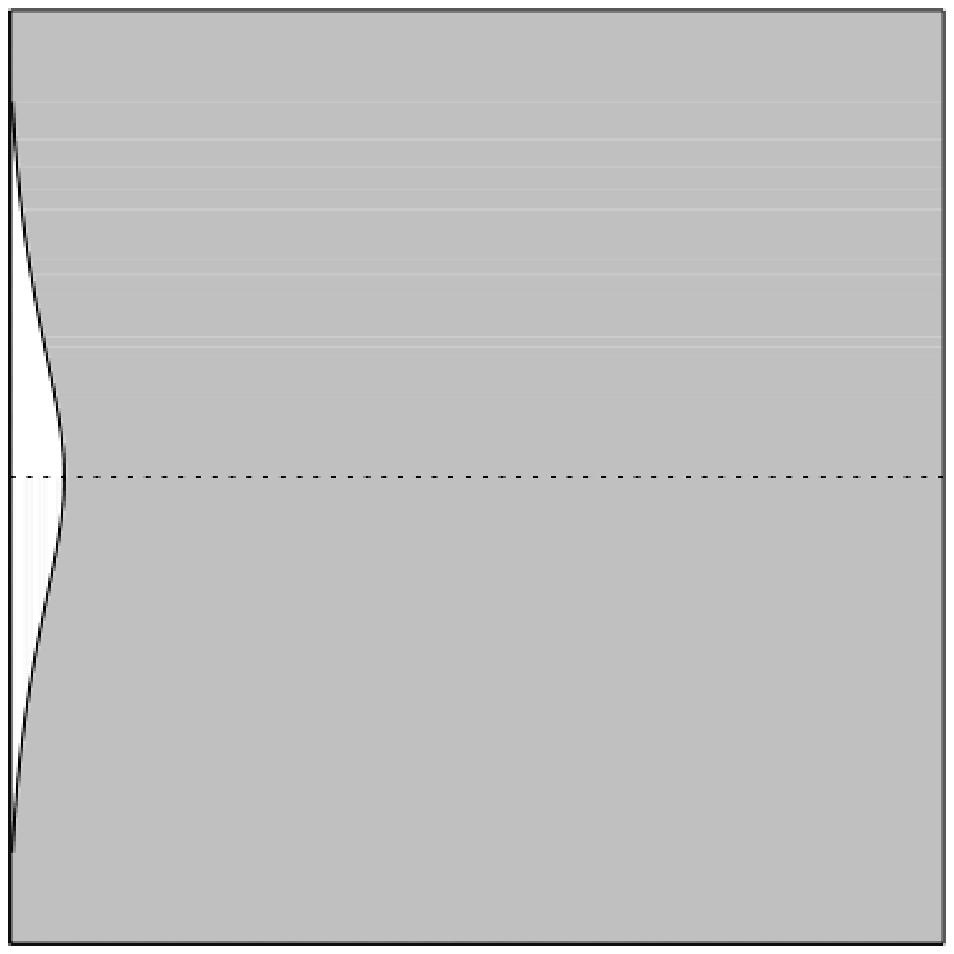}
  \caption{Effective potential for neutral timelike orbits for $m=1$, $E=0.97$ and $\Phi=4$. From left to right, the field strength is $B=0, 0.01$, and $0.03$. The shaded regions indicate areas where $V^2_{\mathrm{eff}}>E^2$ and, hence not accessible to the particles. Horizontal and vertical axes  respectively correspond to $r$ and $\theta$. The horizontal dotted line through the center indicates the equator at $\theta=\pi/2$. The ranges for the above plots are $r\in[2,20]$ and $\theta\in[0,\pi]$.}
  \label{NV_eff_tim}
 \end{center}
\end{figure}

 
Figure \ref{NV_eff_tim} shows the effective potential for neutral timelike particles. The effect is less interesting in this case. The first diagram on the left is the potential for Schwarzschild orbits with $B=0$. Turning on the field strength binds the orbiting particle more closely to the center, hence shrinking the accessible region of the particle.

\subsection{Stability of circular photon orbits}
It is possible to solve the equations of motion to find circular photon orbits \cite{Dhurandhar:1983} where $r$ and $\theta$ are constant. To find such solutions we begin by first observing that for $e=0$, there is a simple solution to the $\theta$ equation with $\theta=\pi/2$. It follows that substituting $\theta=\pi/2$ in \Eqref{ernstN_first} reduces it to a one-dimensional radial potential equation, where circular orbits can be found by considering a value of $r=r_0$ which satisfies
\begin{align}
 \frac{\dif}{\dif r}\sbrac{V^2_{\mathrm{eff}}(r,\pi/2)}=0. \label{circular_N}
\end{align}
Thus it follows from \Eqref{circular_N} and \Eqref{ernstN_first} that the equation for circular, equatorial orbits of radius $r=r_0$ requires the field strength and energy to be given by
\begin{align}
 B^2=\frac{4(r_0-3m)}{r_0^2(3r_0-5m)},\quad E^2=\frac{f(r_0)\Lambda(r_0,\frac{\pi}{2})^2}{r_0^2}\Phi^2. \label{circular_ph_EB}
\end{align}
It appears that $\Phi$ is a free parameter, provided that $E$ and $B$ satisfies \Eqref{circular_ph_EB}. As shown in \cite{Stuchlik:2008xi}, for a given value of $mB$, there are two possible radii in the range $r_0>2m$ which satisfies \Eqref{circular_ph_EB}. The inner radius is unstable while the outer one is stable. The inner and outer radii coalesce at the critical value of 
\begin{align}
 mB=2\beta_{\mathrm{crit}}, \label{B_crit}
\end{align}
where
\begin{align}
 \beta_{\mathrm{crit}}=\sqrt{\frac{3\brac{\sqrt{19}-1}}{\brac{\sqrt{19}+8}^2\brac{\sqrt{19}+3}}}\simeq 0.094\,683\,19\ldots \label{beta_c}
\end{align}
For values satisfying $mB>\beta_{\mathrm{crit}}$, there are no circular orbits. For a given $m$ and $B$ where $mB<2\beta_{\mathrm{crit}}$, there are two circular orbits, where the one with the smaller radius is unstable and the outer one is stable. When $mB$ is tuned to be equal to $2\beta_{\mathrm{crit}}$, the inner and outer orbits coalesce and the circular orbit is marginally stable.  The numerical value of $\beta_{\mathrm{crit}}$ was first obtained in \cite{Dhurandhar:1983} by numerical root-finding, while the analytical expression was obtained later in \cite{Stuchlik:2008xi} by inspecting the condition given in Eq.~\Eqref{circular_ph_EB}.\footnote{To compare the results, it is worth noting that the definition of $B$ in \cite{Dhurandhar:1983} differs from \Eqref{Ernst_metric} by a factor of 2.} We will now demonstrate the stability of the orbits explicitly in a calculation which also takes into account possible motion in the $\theta$ direction. Perturbing around the circular orbit solution, we write
\begin{align}
 r(\tau)=r_0+\varepsilon r_1(\tau),\quad \theta(\tau)=\frac{\pi}{2}+\varepsilon\theta_1(\tau)+\mathcal{O}\brac{\varepsilon^2}, \label{ernstE_circular_ph_perturb}
\end{align}
where $\varepsilon$ is a small perturbation parameter. Substituting \Eqref{ernstE_circular_ph_perturb} into Eqs.~\Eqref{ernstE_rddot} and \Eqref{ernstE_thetaddot} [or \Eqref{ernstM_rddot} and \Eqref{ernstM_thetaddot}] and expanding, we find that the equations for $r$ and $\theta$ decouple at linear order,
\begin{align}
 \frac{\dif^2}{\dif\tau^2}\left(
   \begin{array}{c}
    r_1 \\
    \theta_1
   \end{array}
  \right) =
  \left(
    \begin{array}{cc}
     -\omega^2_r & 0 \\
     0           & -\omega^2_\theta
    \end{array}
  \right)
  \left(
    \begin{array}{c}
      r_1 \\ 
      \theta_1
    \end{array}
  \right), \label{matrix_eqn}
\end{align}
where
\begin{align}
 \omega_r^2=\frac{3r_0^2-16mr_0+15m^2}{2r_0^5(r_0-2m)}\Phi^2,\quad\omega_\theta^2=\frac{m}{r_0^4(r_0-2m)}\Phi^2. \label{ph_frequency}
\end{align}
The stability of the orbits is ensured if both normal mode frequencies $\omega_r$ and $\omega_\theta$ are real. As we can see, $\omega_\theta^2$ is always positive while in the $r$ direction, while $\omega_r^2$ is negative for the range
\begin{align}
 \frac{8-\sqrt{19}}{3}m<2m< r_0\leq \frac{8+\sqrt{19}}{3}m, \label{ph_unstable_range}
\end{align}
where the orbits are unstable. Therefore, circular photon orbits are stable for $r_0>\frac{8+\sqrt{19}}{3}m$. The critical value of $r_0=\frac{8+\sqrt{19}}{3}m$ corresponds with having $mB=\beta_{\mathrm{crit}}$ in Eq.~\Eqref{B_crit}, the value where the inner and outer radii coalesce. Thus, we have demonstrated explicitly that the inner radii are unstable. Furthermore, we have the additional result that the orbits are always stable in the $\theta$ direction. 
\par
\begin{figure}
 \begin{center}
   \begin{subfigure}[b]{0.45\textwidth}
   \centering
   \includegraphics{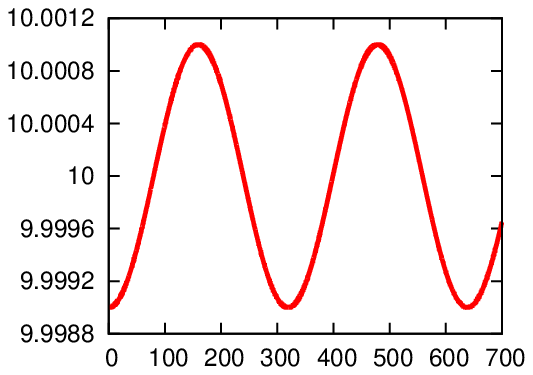}
   \caption{$r=r_0-0.001$, $r_0=10$.}
   \label{fig11a}
  \end{subfigure}
  \begin{subfigure}[b]{0.45\textwidth}
   \centering
   \includegraphics{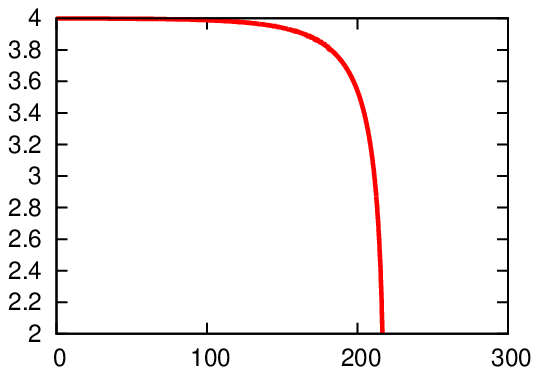}
   \caption{$r=r_0-0.001$, $r_0=4$.}
   \label{fig11b}
  \end{subfigure}
  \end{center}
 \caption{Plots of $r$ vs $\tau$ of perturbed circular photon orbits about (a) $r_0=10$ and (b) $r_0=4$; in both cases, the angular momentum is $\Phi=2$. The equations of motion are solved in units where $m=1$; for case (a) we have $\omega_r^2>0$; hence, it shows a stable oscillation about $r_0=10$. We can check that the numerical solution agrees with the analytical approximation $T=2\pi/\omega_r\simeq 319$. In case (b), $r_0=4$ leads to $\omega_r^2<0$, the perturbed orbit falls into the horizon.}
 \label{fig11}
\end{figure}

 
The calculations can be verified by the numerical solutions. For example, perturbing about a circular photon orbit of radius $r_0=10m$ results in an oscillation with a period of $T\simeq 319$, shown in Fig.~\ref{fig11a} in units where $m=1$. This agrees with $T=2\pi/\omega_r$ where $\omega_r$ is given by \Eqref{ph_frequency}. Fig.~\ref{fig11b} shows an unstable orbit in the range given by \ref{ph_unstable_range}, specifically $r_0=4m$. The instability of such an orbit is clearly seen as the perturbed particle falls beyond the black hole horizon at $r_0=2m$.
\par
\begin{figure}
 \begin{center}
   \includegraphics[scale=1.0]{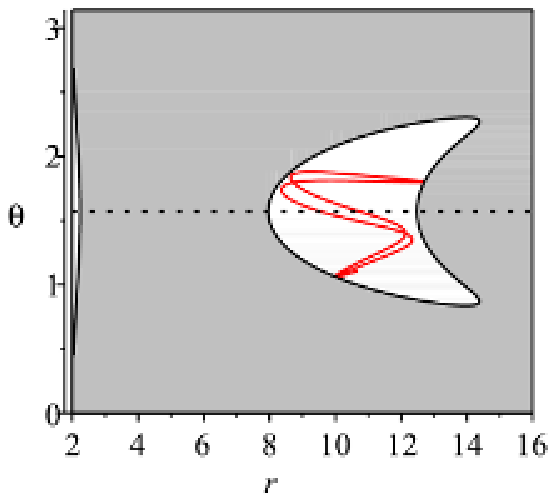}
   \includegraphics[scale=1.0]{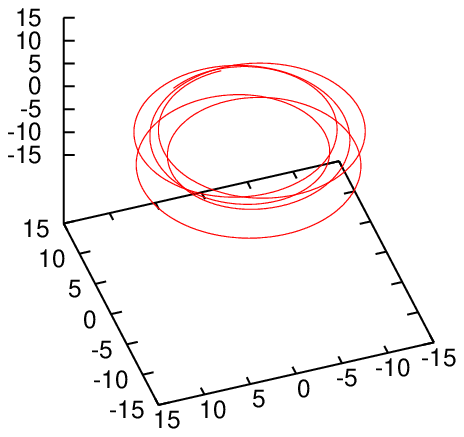}
   \caption{Photon orbit ($\epsilon=e=0$) bound in a finite potential well. The parameters of this solution are $B=0.105$, $E=0.6$, and $\Phi=4$, given in units where $m=1$.}
  \label{N_bound_ph}
 \end{center}
\end{figure}

 
Moving beyond perturbed circular orbits, we find by numerical integration more general bound orbits with nonconstant $r$ and $\theta$. By adjusting $B$, $E$ and $\Phi$, we can find specific parameters where there exists a finite potential well. One such case is shown in the third figure in Fig.~\ref{NV_eff_Ph}. An example of an orbit is shown in Fig.~\ref{N_bound_ph}.

\subsection{Stability of circular timelike orbits}

The perturbation equations for a timelike circular orbit are more complicated than for the photon case. Therefore, we do not have a simple closed-form expression for the stability of orbits analogous to \Eqref{beta_c}, and our result still requires numerical root-finding already considered in \cite{Estaban1984}. 
\par 
Thus, we will only consider this case briefly for completeness. For a given $r_0$ and $B$, circular timelike orbits have energy and angular momentum given by
\begin{align}
 \Phi^2=&\;\frac{16r_0^2\brac{2B^2r_0^3-3B^2mr_0^2-4m}}{\brac{4+B^2r_0^2}^2\brac{5B^2r_0^2m-3B^2r_0^3-12m+4r_0}},\nonumber\\
 E^2=&\;\frac{(r_0-2m)\brac{4+B^2r_0^2}\brac{2B^2r_0^2m-B^2r_0^3-8m+4r_0}}{r_0\brac{5B^2r_0^2m-3B^2r_0^3-12m+4r_0}}.
\end{align}
Perturbing around the orbits in the case of timelike orbits also yields decoupled equations of the form \Eqref{matrix_eqn}, but with the frequencies given by
\begin{align}
 \omega_r^2=&\;\frac{16}{r_0^3\brac{4+B^2r_0^2}^4(5B^2mr_0^2-3B^2r_0^3-12m+4r_0)}\bigl[(12r_0^8-37mr_0^7+30m^2r_0^6)B^6\nonumber\\
      &\hspace{2cm}+(204mr_0^5-200m^2r_0^4-48r_0^6)B^4+(672m^2r_0^2+128r_0^4\nonumber\\
      &\hspace{2cm}-624r_0^3m)B^2+64mr_0-384m^2\bigr]\\
 \omega_\theta^2=&\;\frac{16m\brac{4-B^2r_0^2}}{r_0^2\brac{4+B^2r_0^2}^2\brac{5B^2r_0^2-3B^2r_0^3-12m+4r_0}}.
\end{align}
Stable orbits require $\omega_r^2>0$ and $\omega_\theta^2>0$; the value of $r_0$ required for stability differs for different $B$, unlike in the photon case.

\section{Conclusion} \label{conclusion}

In this paper we have considered the geodesic equations for charged particles in the Ernst metric. The metric represents a Schwarzschild black hole immersed in an axisymmetric electric or magnetic field. In the case of the electric Ernst metric, the geodesic equations describe charged particles experiencing a central force in addition to a uniform constant force along the axial direction. We find that nearly circular polar and equatorial orbits of radius $r_0>6m$ are stable in weak electric fields, in accordance with the stability of Schwarzschild circular orbits. In the case of polar orbits, the radial equation reduces to that of a harmonic oscillator driven by a periodic driving force. This driving force corresponds to the periodic frequency of the motion in the $\theta$ direction. When the field strength is not necessarily small, it is still possible to have circular trajectories whose orbital plane is parallel to the equatorial plane but lies at a fixed distance above it. 
\par 
For the case where the Maxwell field is purely magnetic, we considered curly trajectories in the Melvin spacetime. Such orbits may form if the charged particle simultaneously experiences an inward gravitational force which is counteracted by a Lorentz force directed outwards. Such motion exists already in the domain of classical electromagnetism for particles in uniform crossed electric and magnetic fields \cite{Jackson:2007}. For particles in the Melvin spacetime, instead of an electric Coulomb force we have the geodesic motion in a spacetime curved magnetic field itself.
\par 
We have also considered orbits in weak magnetic fields by a perturbative expansion of the full equations of motion in the Ernst spacetime. With this expansion, we recover the equations of motion in the Wald's construction of weakly magnetized black holes, where the magnetic field is a test field that does not influence the spacetime curvature. By taking higher-order terms of the expansion, we are able to calculate gravitational corrections to the results of the test field case.
\par 
Neutral particles do not distinguish between the magnetic and electric nature of the fields; hence, the equations of motion for $e=0$ for the electric and magnetic cases reduce into each other identically. This case is studied in \cite{Karas1990} and \cite{Estaban1984} by focusing on motion confined on the equatorial plane. Here we have shown explicitly that photon orbits of radius $r_0>\frac{8+\sqrt{19}}{3}m$ are indeed stable even when the particles are perturbed slightly away from the equator.
\par 
In this paper we have focused exclusively on electrically charged and neutral timelike particles, in addition to (neutral) photons. We have not explicitly considered more exotic particles such as magnetic monopoles or charged massless particles. Theoretically, the motion of magnetic monopoles in the magnetic Ernst spacetime should be identical to electric monopoles in the electric Ernst spacetime. So we should expect most of the results obtained already for electrically charged particles should carry over after performing the appropriate duality operations. The motion for charged massless particles is contained in the equations of motion in Sec.~\ref{ernstE} and \ref{ernstM}, by considering $e\neq0$ and $\epsilon=0$.

\section*{Acknowledgements}
The author would like to thank Edward Teo for illuminating discussions and comments.

\end{document}